\documentclass[reprint,amsmath,amssymb,aps,prb,floatfix,referee]{revtex4-1}
\usepackage{graphicx}
\usepackage{dcolumn}
\usepackage{bm}
\usepackage{multirow}
\usepackage{color}
\usepackage[export]{adjustbox}
\usepackage[caption=false,position=top]{subfig}
\usepackage[english]{babel}
\usepackage{placeins}
\usepackage[T1]{fontenc}
\usepackage{amsmath}
\usepackage{amsfonts}
\usepackage{amssymb}
\usepackage{natmove}
\usepackage{color}
\usepackage{tabularx,booktabs}
\usepackage[sort&compress]{natbib}
\usepackage{tikz}
\usetikzlibrary{shapes}
\newcolumntype{Y}{>{\centering\arraybackslash}X}
\usepackage{dcolumn}
\usepackage[unicode=true, bookmarks=true, bookmarksnumbered=false, bookmarksopen=false,
  breaklinks=false,pdfborder={0 0 0},backref=false,colorlinks=true]{hyperref}
 \hypersetup{pdftitle={Pressure-Induced Topological Phase Transition in CdGeSb$_2$ and CdSnSb$_2$},
  pdfauthor={Rinkle Juneja, Ravindra Shinde and Abhishek K. Singh},
  pdfsubject={Computational Material Science},
  unicode=false,pdftoolbar=true,pdfmenubar=true,pdffitwindow=true,pdfstartview={FitH},pdfnewwindow=true,pdfcreator={RevTeX},linkcolor=red,citecolor=blue,urlcolor=blue}

\newcommand{\markercircle}{\raisebox{0.5pt}{\tikz{\node[draw,scale=0.4,circle,fill=red,red,opacity=0.9](){};}}}
\newcommand{\markersquare}{\raisebox{0.5pt}{\tikz{\node[draw,scale=0.4,regular polygon, regular polygon sides=4,fill=blue,blue,opacity=0.9](){};}}}
\newcommand{\markerdiamond}{\raisebox{0pt}{\tikz{\node[draw,scale=0.4,diamond,fill=black!60!green,black!60!green](){};}}}
    
\newcommand{\markertriangle}{\raisebox{0.5pt}{\tikz{\node[draw,scale=0.3,regular polygon, regular polygon sides=3,fill=orange,orange, opacity=0.9, rotate=0](){};}}}
 
\begin{document}

\title{Pressure-Induced Topological Phase Transitions in CdGeSb$_2$ and CdSnSb$_2$} 
\author{Rinkle Juneja}
\thanks{Equal contribution}
\author{Ravindra Shinde}
\thanks{Equal contribution}
\author{Abhishek K. Singh}
\email{abhishek@iisc.ac.in}
\affiliation{Materials Research Center, Indian Institute of Science, Bangalore, Karnataka 560012, India}
\date{\today}

\begin{abstract}
Topological quantum phase transitions (TQPTs) in a material induced by external perturbations are often characterized by band touching points in the Brillouin zone. The low-energy excitations near the degenerate band touching points host different types of fermions, while preserving the topological protection of surface states. 
An interplay of different tunable topological phases offers an insight into the evolution of topological character.
 In this paper, we study the occurrence of TQPTs as a function of hydrostatic pressure in CdGeSb$_2$ and CdSnSb$_2$ chalcopyrites, using the first-principles calculations.  At ambient pressure, both materials are topological insulators having a finite band gap with inverted order of Sb-$s$ and Sb-$p_x$,$p_y$ orbitals of valence bands at the $\Gamma$ point. On the application of hydrostatic pressure the band gap reduces, and at the critical point of the phase transition, these materials turn into Dirac semimetals.  On further increasing the pressure beyond the critical point, the band inversion is reverted making them trivial insulators. The pressure-induced change in band topology from non-trivial to trivial phase is also captured by L\"{u}ttinger model Hamiltonian calculations. Our model demonstrates the critical role played by a pressure-induced anisotropy in frontier bands in driving the phase transitions. These theoretical findings of peculiar coexistence of multiple topological phases in the same material provide a realistic and promising platform for the experimental realization of the TQPT.
\end{abstract}

\maketitle
\section{Introduction}

Topological insulators (TIs) have triggered a surge of research activities due to their peculiar properties.  This intriguing state of matter is characterized by non-zero topological invariants and exhibits unique features such as an insulating gap in the bulk and robust metallic surface or edge Dirac states protected by time-reversal symmetry \cite{2005Prl-Kane-Z2,2007Prl-Fu-TI3D,2010ReviewMP-Hasan-colloquium}. Till now numerous two- and three-dimensional TIs have been discovered with the spin-orbit coupling induced parity inversion of conduction and valence bands as a guiding principle\cite{2006science-Bernevig-HgTe,2008nature-Hasan-DSM,2008PRB-Kane-BiSb,2009naturephysics-Zhang-TIs,2009science-Chen-Bi2Te3,2014PRB-Zhong-superconductorPbSnTe,2011ReviewMP-Qi-TIs}. 

Dirac semimetals (DSM), on the other hand, have metallic states touching only at a single point near the Fermi energy with a linear band dispersion\cite{2014science-Liu-diracsemimetalNa3Bi,2014naturecommun-Neupane-diracsemimetalCd3As2, 2012PRB-Wang-A3Bi-dirac, 2014Naturecommun-Yang-diracsemimetal,thsim-sunny,thsim-ranjan}. The bulk band gap of topological insulators too can be closed, thereby placing the topologically protected metallic bulk states at the critical point of a topological quantum phase transition (TQPT). This has been realized by tuning an intrinsic materials' property such as spin-orbit interaction by alloying composition or chemical doping  \cite{2011-Science-hasan-texture-inversion, 2012Prl-Brahlek-BiInSe-TI-transition, 2011-NatPhy-Sato-QPT,2013-NatPhy-Wu-TPT, 2014-PRL-Yan-doping-transition}. For example, a quantum critical transition from a topological insulator to a trivial insulator has been observed in TlBiSe$_{2-x}$S$_x$ \cite{2015PRB-Novak-TlBiSSe,2012PRB-Singh-TlBiSe2} and Hg$_{1-x}$Cd$_x$Te \cite{2014Naturephysics-Orlita-3Dkane-fermions}, where a single Dirac point occurs at the time-reversal-invariant-momenta.  The transition in these compounds is very sharp in terms of exact alloying concentration, and realizing an accurate, stable, homogeneous 3D Dirac material in these systems is an experimental challenge\cite{2014Naturephysics-Orlita-3Dkane-fermions,2012Prl-Brahlek-BiInSe-TI-transition, 2013-NatPhy-Wu-TPT}. 

These transitions can also be achieved through pressure without the need of any doping or alloying. Besides, this method circumvents the problem of unwanted defects and inhomogeneity of doping. Hence, it has become a popular tool for observing quantum phase transitions in topological materials. For example, strain-induced topological phase transition was studied for elemental tellurium\cite{2013-PRL-goddard-tellurium} and HgSe\cite{2013-PRB-Chen-strain}. Pressure-induced topological phase transition in rocksalt chalcogenides\cite{2013-PRB-Barone-pressure}, layered materials\cite{2012-PRL-KAUST-pressure-layered}, BiTeI\cite{2013-PRL-Carr-TQPT}, Pb$_{1-x}$Sn$_{x}$Se\cite{2014-PRL-Carr-bulk-signatures}, bialkali bismuthide KNa$_2$Bi\cite{2016scireports-Sklydneva-KNa2Bi} and polar semiconductors BiTeBr\cite{2017-PRB-japan-BiTeBr}, BiTeI\cite{2015Scireports-Park-BiTeI} was observed, while topological insulator phase in NaBaBi\cite{2016-PRB-Sun-pressure} and Zintl compounds\cite{2016-PRB-Zintl} was obtained starting from a semimetallic state. The phase transition due to pressure was also observed in topological crystalline insulators \cite{2015-PRB-Zhao-TCI}. 

To observe the topological phase transition by detecting the gapless metallic surface states, surface-sensitive experimental techniques such as angle-resolved photoemission spectroscopy (ARPES), scanning tunneling microscopy are used. However, performing ARPES to observe surface states at high pressures is a challenging task; making it experimentally difficult to study pressure-induced phase transitions in topological materials.  On the other hand, the theoretical investigations provide an alternative feasible way to probe these transitions, as well as to characterize simultaneously the peculiar surface states before performing any experiments. Hence, the interest in the theoretical exploration of pressure-induced TQPTs has been renewed. 

In this paper, we report pressure-induced topological phase transitions in chalcopyrite compounds CdGeSb$_2$ and CdSnSb$_2$ by first-principles density functional based approach. These materials are small band gap topological insulators in their equilibrium states. The bulk band gap of these materials decreases upon the application of pressure, and materials transform to a DSM at a critical pressure. The band gap reopens with a normal band order above the critical pressure. We also simulate the ARPES at the ambient pressure as well as at the critical point of TQPT using an \textit{ab initio} tight-binding calculations. These TQPTs are further substantiated by a generalized model Hamiltonian calculations, which emulates the evolution of bandstructure with the application of pressure. This model shows the breaking of spherical isotropic nature of bands and the emerging dominance of anisotropic terms as a function of pressure, thereby supporting the non-trivial to trivial phase transition.  These theoretical investigations provide an ideal platform to study multi-topological phase transitions in the same material and may serve as a guide for experimental realization.


\captionsetup[subfigure]{font=normal, labelfont=normal,  labelfont=bf, textfont=normalfont}

\begin{figure}
\centering
\includegraphics[width=\columnwidth]{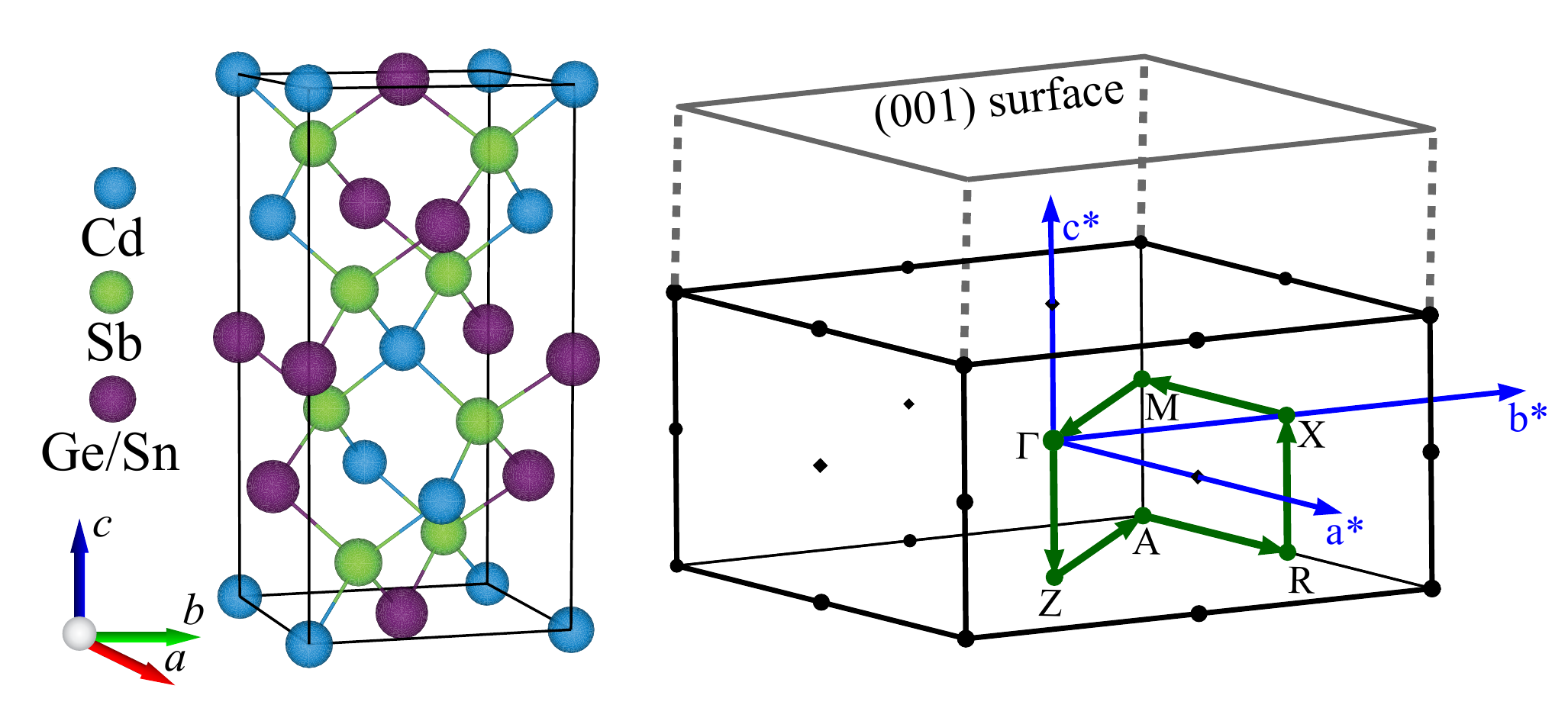}
\vspace{-6mm}
\caption{(color online). The conventional unit cell of II-IV-V$_2$ chalcopyrite compounds CdGeSb$_2$/CdSnSb$_2$ with corresponding Brillouin zone and its projection onto (001) surface along with high-symmetry points.}
\label{fig:crystal-structure}
\end{figure}

\section{Methodology}
Theoretical calculations were performed within the first-principles density functional theory (DFT) \cite{1965-PhysRev-Kohn-Vxc}  using Vienna Ab initio Simulation Package (VASP) \cite{1966CMS-Kresse-Ecalcs,*1996PRB-Kresse-iterativeEcalcs}. Core electrons were represented by projector-augmented wave potentials \cite{1994PRB-Blochl-PAW,*1999PRB-Kresse-PAW}. Since PBE functional underestimates the band gap, it may not be applicable in the case of small band gap materials with strong spin-orbit coupling, particularly where materials are classified as topological insulators with a false band inversion.  Hence we used a hybrid HSE functional which takes 25\% of the short-range exact exchange. \cite{HSE-2003, *HSE-2006}

\begin{figure}[t!]
\subfloat[\label{subfig:Sn-bands}CdSnSb$_2$ w/o SOC \hspace{1cm} \textbf{(b)} with SOC]{\includegraphics[width=\columnwidth]{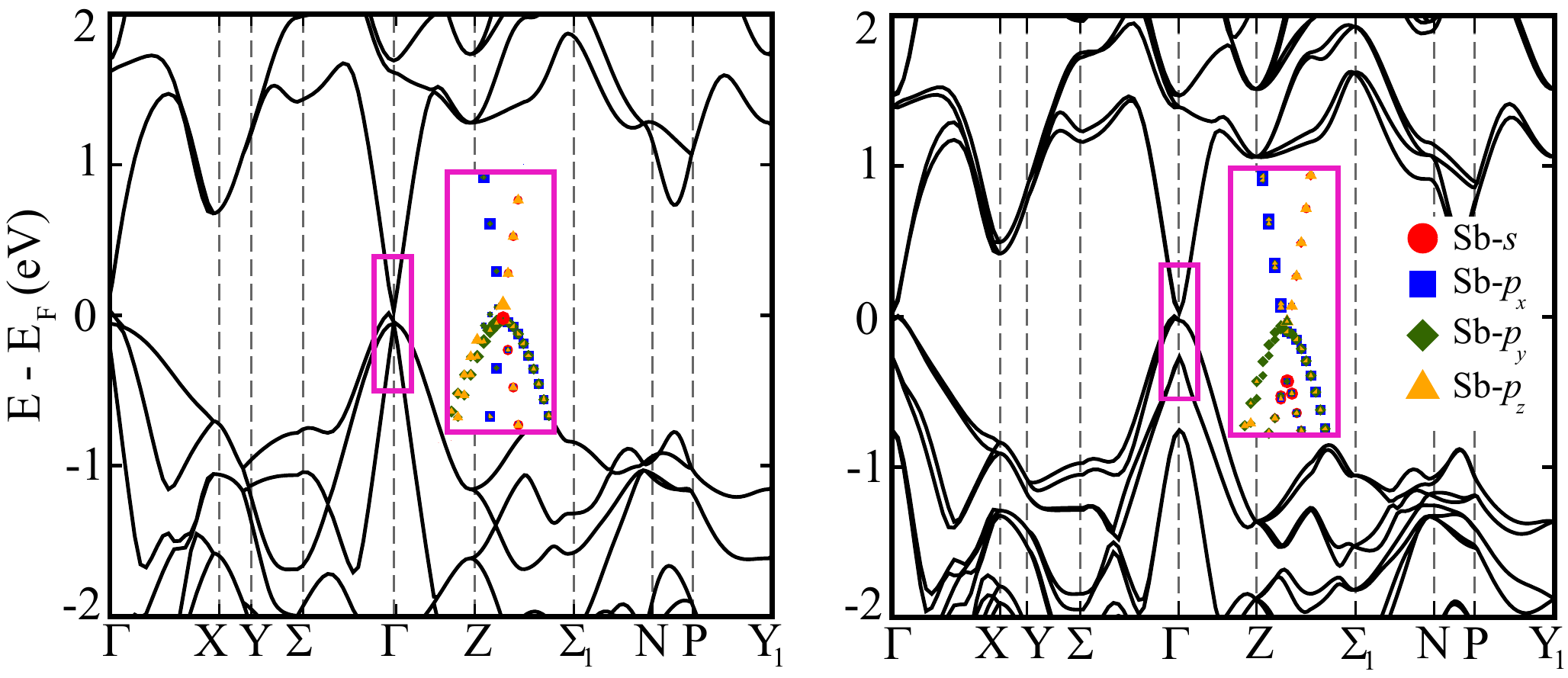}} \\
\setcounter{subfigure}{2}
\subfloat[\label{subfig:Ge-bands}CdGeSb$_2$ w/o SOC \hspace{1cm} \textbf{(d)} with SOC]{\includegraphics[width=\columnwidth]{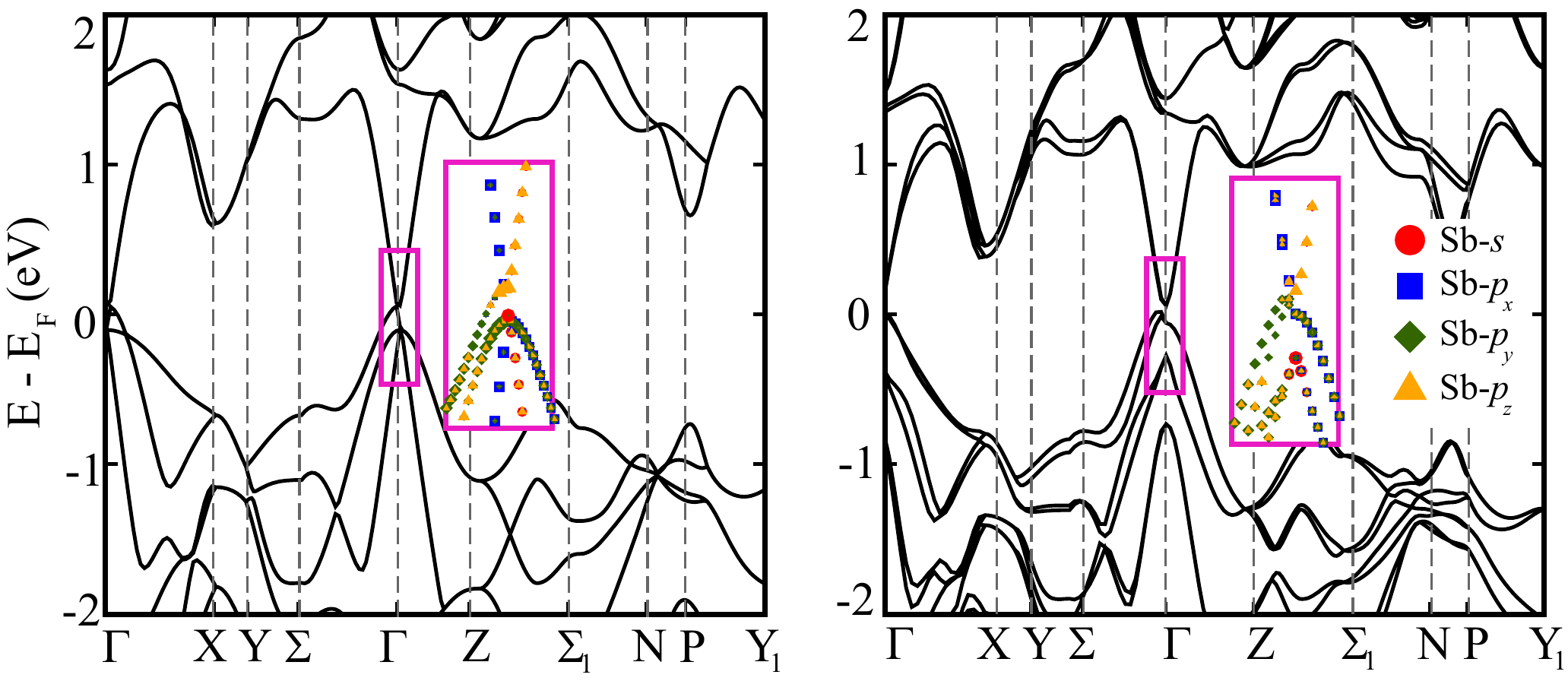}} \\
\vspace{-2mm}
\caption{(color online). Orbital-resolved electronic band structure of CdSnSb$_2$ and CdGeSb$_2$ without spin-orbit coupling \protect\subref{subfig:Sn-bands} and \protect\subref{subfig:Ge-bands} and with spin-orbit coupling (b) and (d). The orbital contribution of atoms is denoted by {$s$ -- \protect\markercircle}, $p_x$ -- {\protect \markersquare}, $p_y$ -- {\protect \markerdiamond} and $p_z$ -- {\protect \markertriangle}.}
\label{fig:band-structure}
\end{figure}

Relativistic effects were included in the calculations with a plane-wave basis energy cut-off of 400 eV and a $\Gamma$-centered Monkhorst-Pack \cite{1976PRB-Monkhorst-BZintegration} \textbf{k}-grid of 5$\times$5$\times$5. All the structures, after the application of hydrostatic pressure, were relaxed by employing a conjugate-gradient scheme until the forces on each atom become less than 0.005 eV/\AA{}. The band structures were then calculated with and without SOC using these optimized structures.  

The topological invariance, characterized by $\mathbb{Z}_{2}$ number, was calculated by the method of evolution of Wannier charge centers (WCC), as implemented in the Z2Pack.\cite{ 2011-PRB-Soluyanov-Z2} The charge centers of a given set of Wannier functions are defined as the average position of charge of a Wannier function in the chosen unit cell. The evolution of WCC along any direction in $k$-space corresponds to the change in the phase factor $\theta$ of the eigenvalues of the position operator projected onto the occupied subspace.\cite{2006-PRB-Kane-Z2, 2011-PRB-Soluyanov-Z2, 2011-PRB-Bernevig-Z2} The odd number of crossings of any random horizontal reference line with the evolution of $\theta$'s signifies a topological insulator with $\mathbb{Z}_{2}$ = 1. 
A tight-binding model was built based on maximally localized Wannier functions (MLWF) obtained using \textit{ab initio} DFT results. An iterative Green's function method was employed to obtain the surface density of states \cite{2014ComputPhysCommun-Mostofi-Wannier90,opensource-QuanSheng-Wannier-tools}. A generalized L\"{u}ttinger $j=3/2$ model Hamiltonian is constructed to mimic the evolution of HSE bands with pressure.

\section{Results and Discussion}
\subsection{Crystallographic structure and electronic properties}
The compounds studied here belong to the class of II-IV-V$_2$ materials which crystallize in the tetragonal chalcopyrite structure with I-42d space group \cite{tengaa2005,shay2013}. In this structure, each of the group II and group IV elements are tetrahedrally coordinated by four group V elements while the group V elements are tetrahedrally coordinated by two group II and two group IV elements, as shown in Fig. \ref{fig:crystal-structure}. Since the group V atom is bonded to two different types of cations, the respective bond lengths are not necessarily the same. Hence the c/a ratio deviates from an ideal value of 2. The optimized lattice parameters for the CdSnSb$_2$ and CdGeSb$2$ unit cells are a=b=6.63 \AA{}, c=13.11 \AA{}, and a=b=6.45 \AA{}, c=12.48 \AA{}, respectively. The corresponding Brillouin zone with the high-symmetry points is also shown in Fig. \ref{fig:crystal-structure}.

\begin{figure}
\centering
\subfloat{\raisebox{-2.8cm}{\rotatebox[origin=t]{90}{E -- E$_F$ (eV)}}} 
\setcounter{subfigure}{0}
\subfloat[]{\includegraphics[trim={0 0 0.58cm 0},clip, width=0.415\columnwidth]{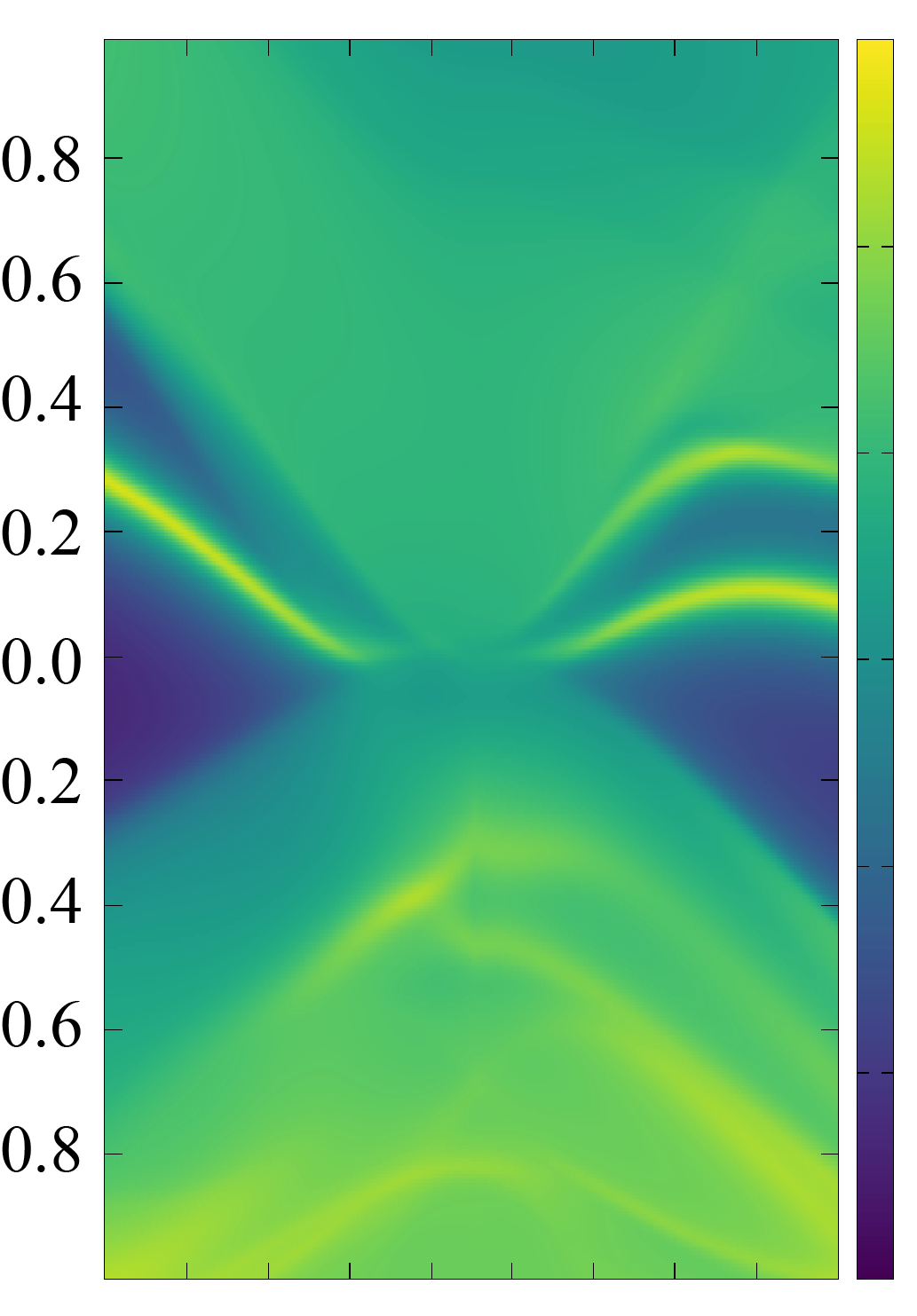}} \hspace{2mm}  %
\subfloat[]{\includegraphics[trim={0.8cm 0 0 0},clip, width=0.41\columnwidth]{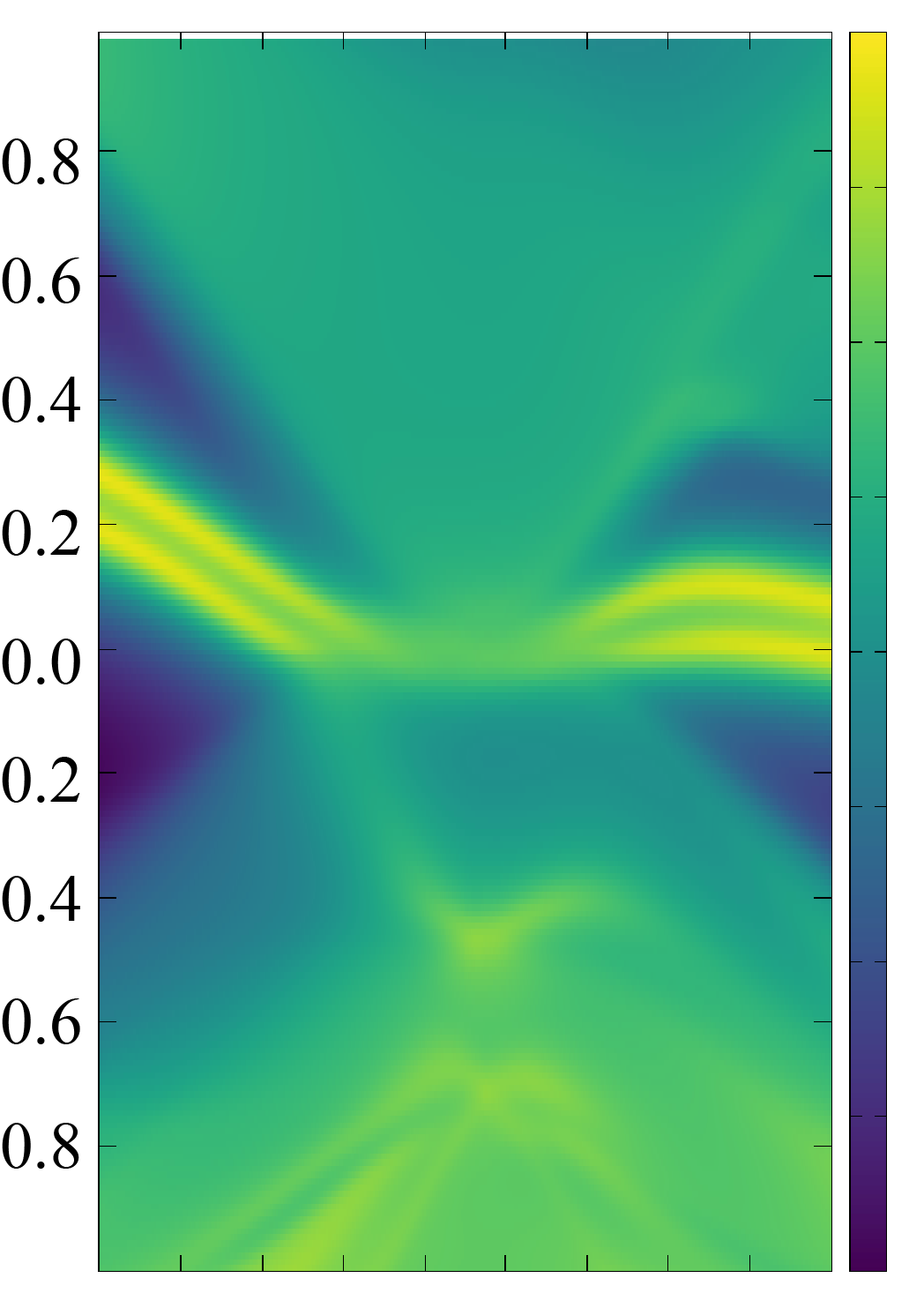}}                %
\vspace{-3mm}
\caption{\label{fig:arpes_and_fermi-ambient} (color online) Momentum resolved (001) surface density of states calculated near $\Gamma$ point for (a) CdSnSb$_{2}$ and (b) CdGeSb$_{2}$ with SOC in the equilibrium phase.}
\end{figure}

Figure \ref{fig:band-structure} shows the HSE band structures of CdSnSb$_2$ and CdGeSb$_2$ without and with spin-orbit coupling (SOC). The inset shows the orbital contribution of Sb atoms, which has major contribution of orbitals in the vicinity of $\Gamma$ point for both these compounds. In the absence of SOC, the valence band maxima (VBM) has the major contribution from Sb-$s$ orbitals, followed by the bands with Sb-$p_{x}$ and Sb-$p_{y}$ contribution, whereas the conduction band minima (CBM) has Sb-$p_{z}$ character dominating, along with minor Sb-$p_{x}$ and Sb-$p_{y}$ contribution. On the inclusion of SOC, a small gap of 0.021 and 0.058 eV is opened for CdSnSb$_2$ and CdGeSb$_2$, respectively. Furthermore, the order of valence bands gets inverted at the $\Gamma$ point due to interchange of Sb-$s$ and Sb-$p_{x}$, Sb-$p_{y}$ character, for both the compounds. Hence, CdSnSb$_2$ and CdGeSb$_2$ are non-trivial topological insulators.

\begin{figure}[t!]
\centering
\subfloat{\raisebox{-2.8cm}{\rotatebox[origin=t]{90}{E -- E$_F$ (eV)}}} 
\setcounter{subfigure}{0}
\subfloat[]{\includegraphics[trim={0 0 0.7cm 0},clip, width=0.417\columnwidth]{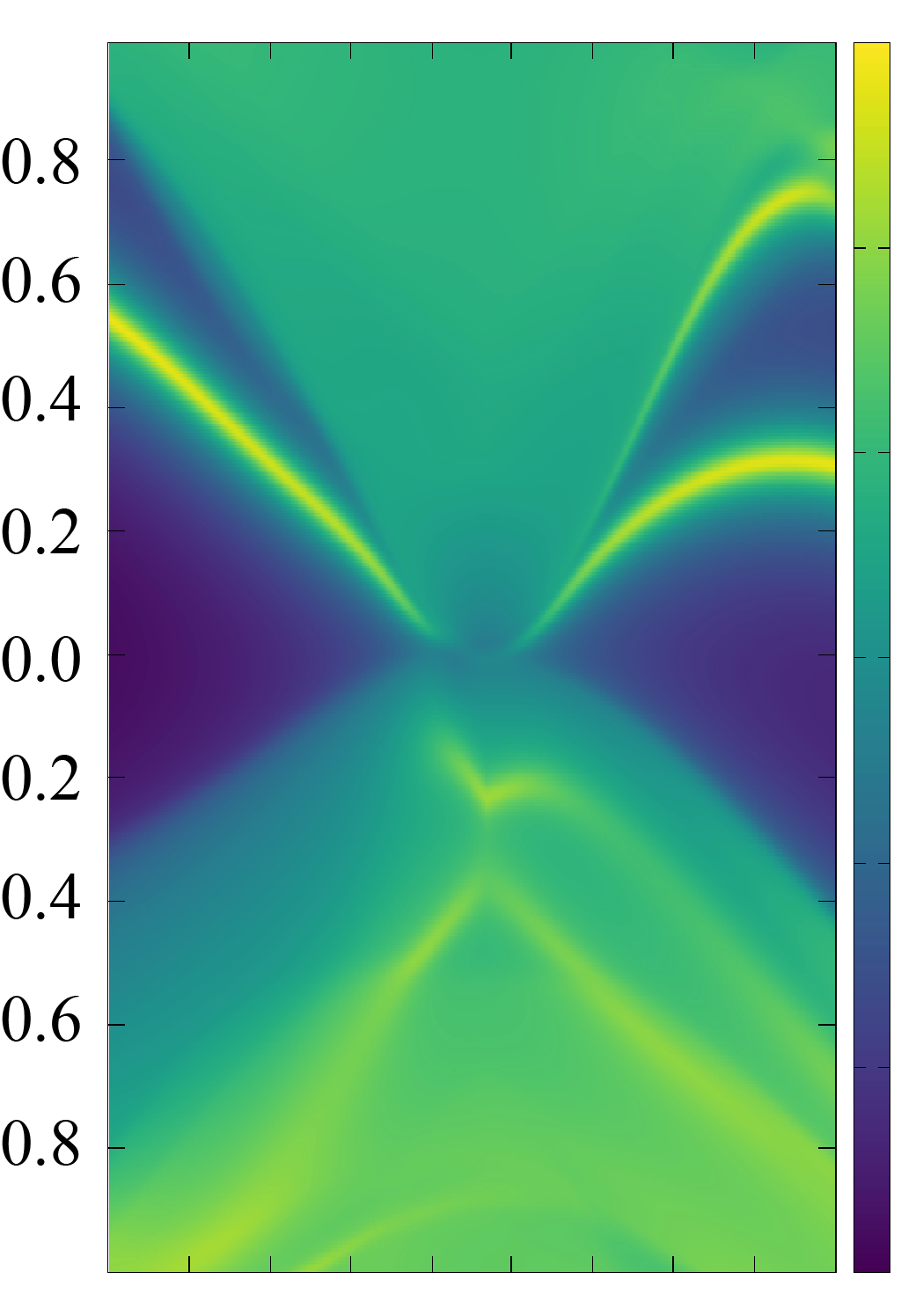}} \hspace{2mm}
\subfloat[]{\includegraphics[trim={0.8cm 0 0 0},clip, width=0.41\columnwidth]{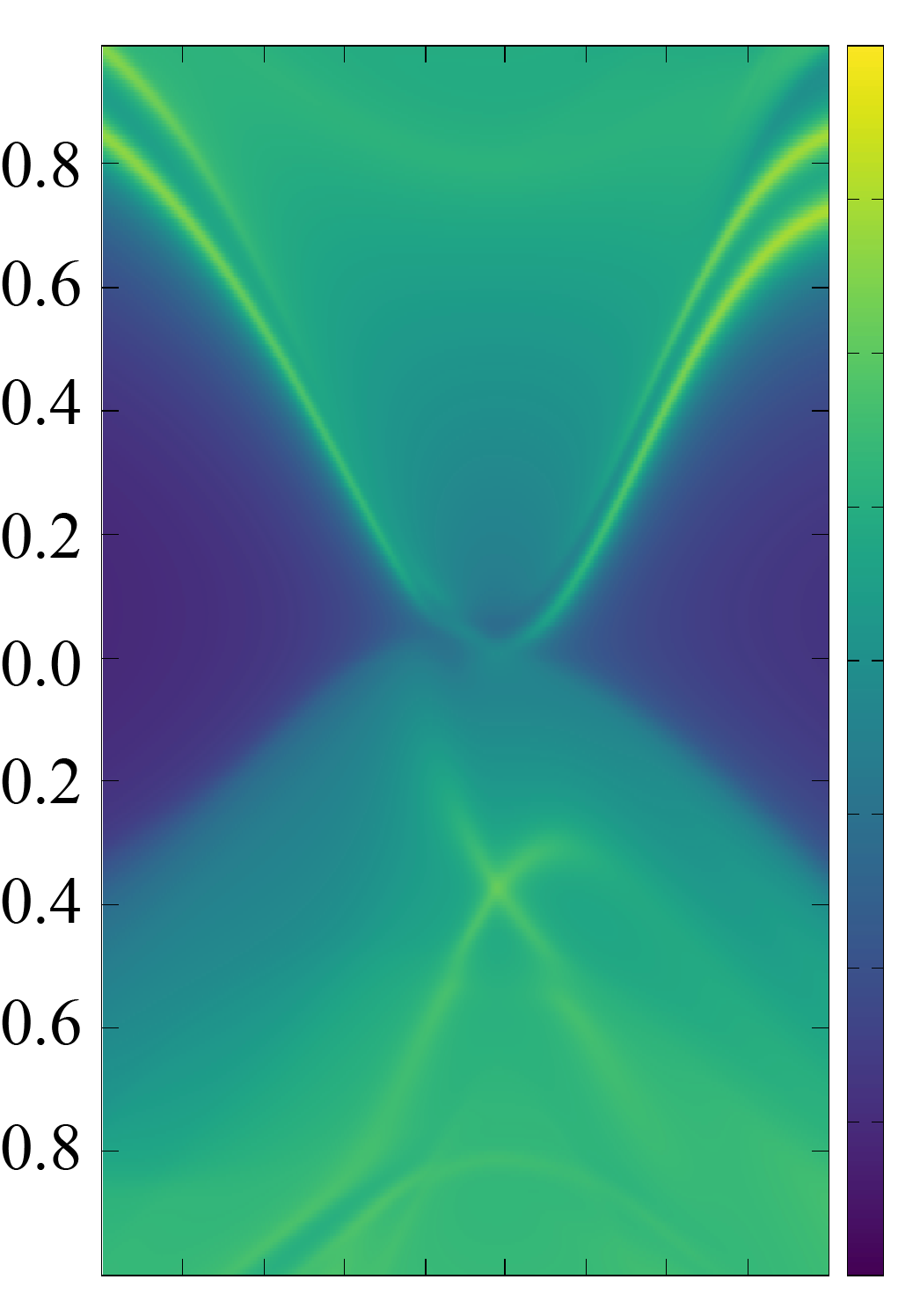}}
\vspace{-3mm}
\caption{\label{fig:arpes_and_fermi-Dirac} (color online) surface density of states for (a) CdSnSb$_{2}$ and (b) CdGeSb$_{2}$ with SOC at the critical point under hydrostatic pressure. Surface states are seen connecting at the Dirac point.}
\end{figure}

\begin{figure}[b]
\centering
\subfloat{\raisebox{-1.4cm}{\rotatebox[origin=t]{90}{Energy}}} 
\setcounter{subfigure}{0}
\subfloat[]{\includegraphics[width=0.4\columnwidth]{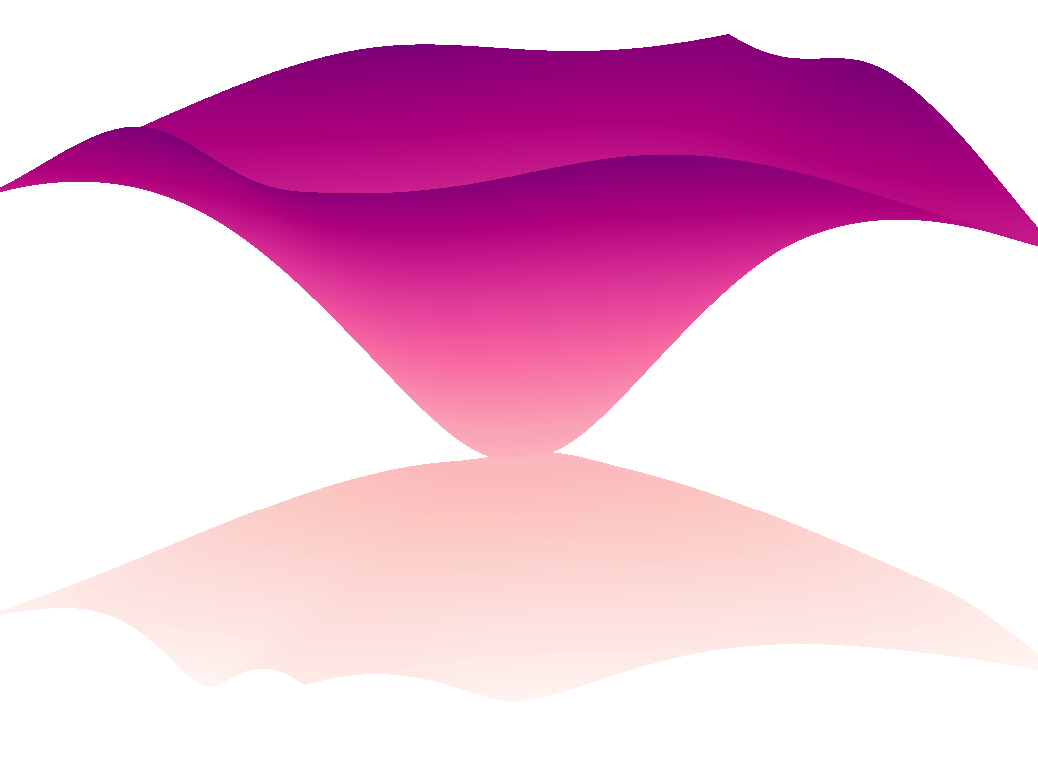}}
\subfloat[]{\includegraphics[width=0.4\columnwidth]{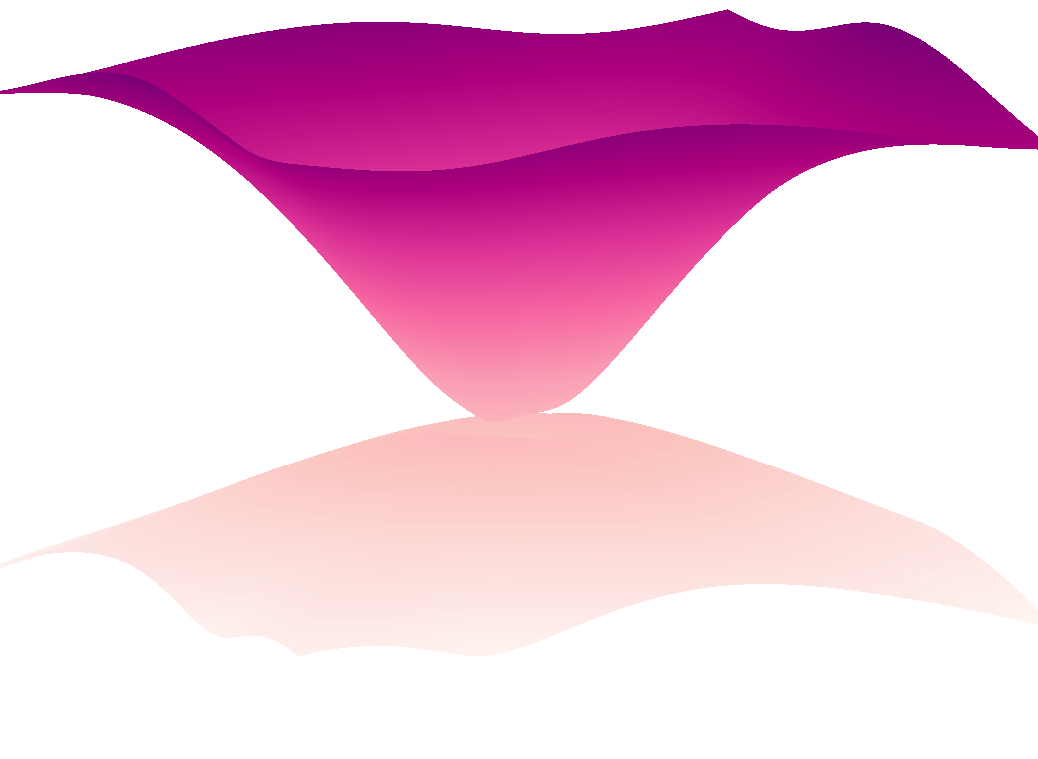}}
\vspace{-3mm}
\caption{\label{fig:dirac_cone} (color online) The two-dimensional energy dispersion of at the critical point (a) CdSnSb$_2$ and (b) CdGeSb$_2$.}
\end{figure}

Since the non-trivial topological character should be reflected on the surface, we did Z-terminated (001) plane surface state calculations for CdSnSb$_2$ and CdGeSb$_2$, based on the idea of the bulk-edge correspondence of the TIs. \cite{1994PRB-Blochl-PAW} The surface density of states with SOC, calculated using iterative Green's function method are shown in Fig. \ref{fig:arpes_and_fermi-ambient} and Fig. \ref{fig:arpes_and_fermi-Dirac} for ambient pressure phase and at critical pressure, respectively. The tight-binding model based on the maximally-localized Wannier functions correctly reproduces DFT bandstructure and simulates the ARPES with the calculated surface density of states. The energy dispersion in two dimensions at the critical pressure for both the materials is shown in Fig. \ref{fig:dirac_cone}. To further confirm the topological nature for both the compounds, we also calculated $\mathbb{Z}_{2}$ topological invariant through the evolution of Wannier charge centers. The $\mathbb{Z}_{2}$ for these compounds comes out to be one, thereby substantiating their non-trivial topology.

Authors of Refs. \onlinecite{2011-feng-prl-controversial, 2012-Feng-Chinese-Review} have also predicted these materials to be topological insulators in their equilibrium state with slightly larger bandgaps. This disparity might have arisen because the authors\cite{2011-feng-prl-controversial, 2012-Feng-Chinese-Review} have taken an ideal chalcopyrite structure with c/a = 2.0; thereby overestimating the strain. 

  
 \begin{figure*}[ht!]
\subfloat[\label{subfig:Sn-pressure}CdSnSb$_2$]{\includegraphics[width=\textwidth]{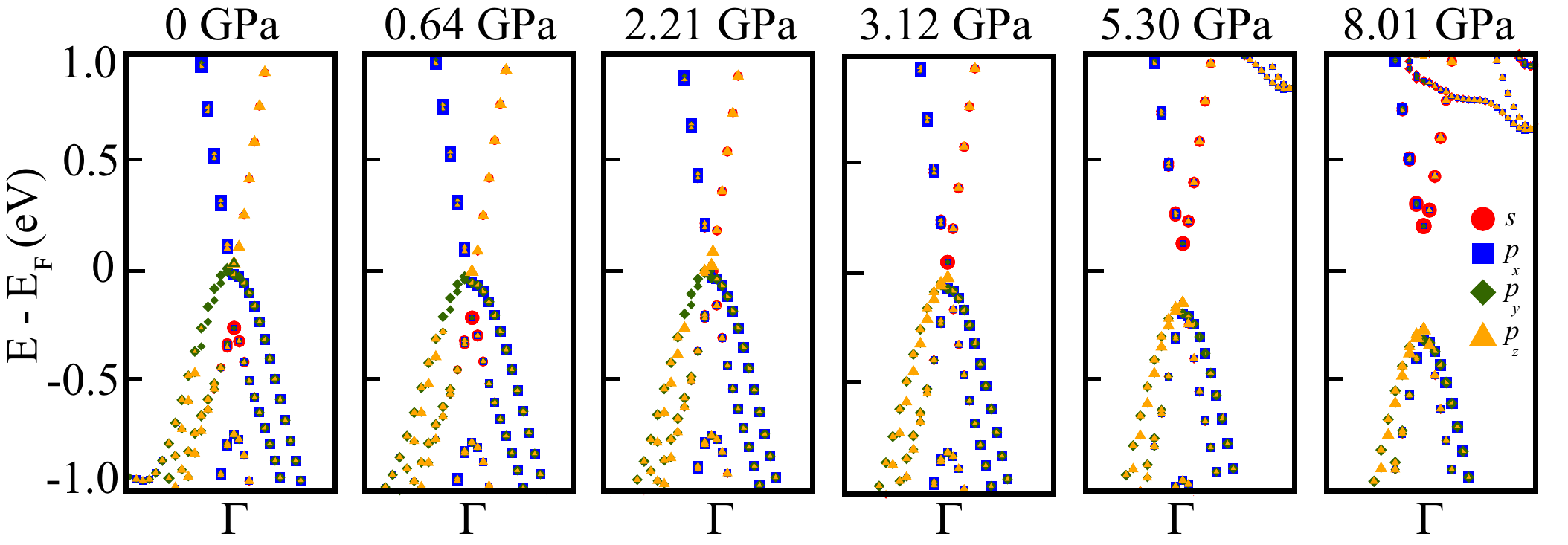}} \\
\subfloat[\label{subfig:Ge-pressure}CdGeSb$_2$]{\includegraphics[width=\textwidth]{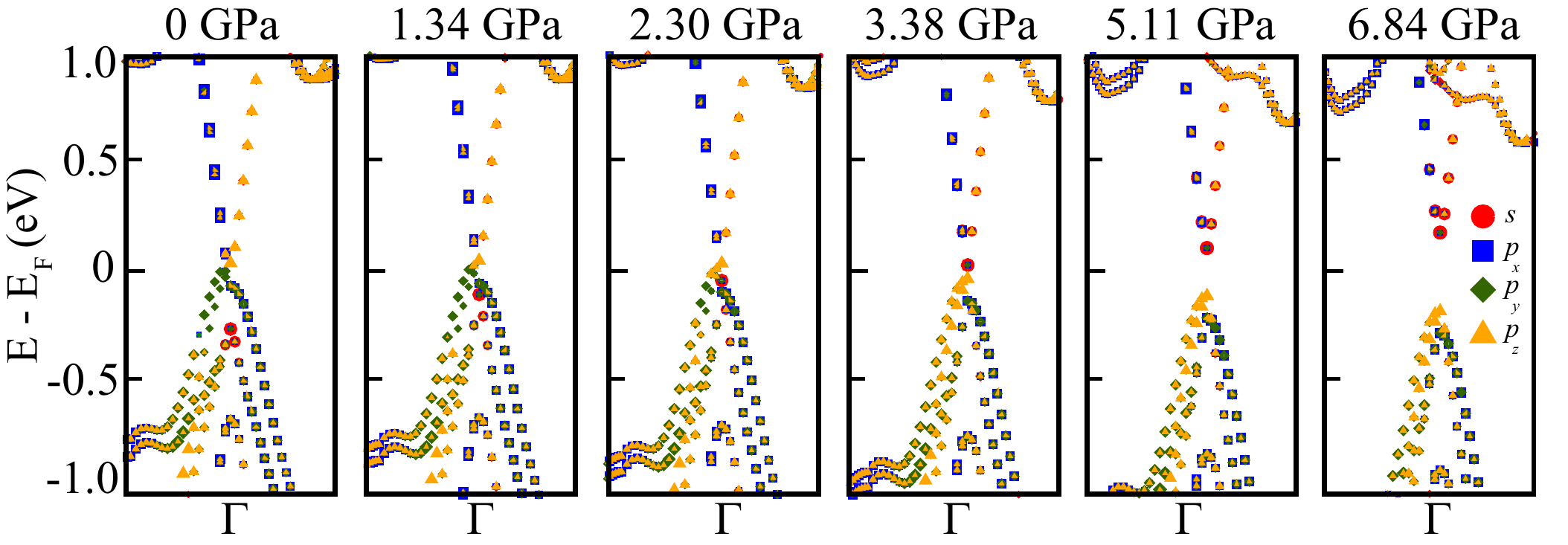}} \\
\caption{Effect of hydrostatic pressure on the band structure of \protect\subref{subfig:Sn-pressure} CdSnSb$_2$ and \protect\subref{subfig:Ge-pressure} CdGeSb$_2$. The orbital contribution of atoms is denoted by {$s$ -- \protect\markercircle}, $p_x$ -- {\protect \markersquare}, $p_y$ -- {\protect \markerdiamond} and $p_z$ -- {\protect \markertriangle}.}
\label{fig:pressure-bands}
\end{figure*}
 
 
\subsection{Effect of hydrostatic pressure} Next, we analyzed how the electronic structures of CdSnSb$_2$ and CdGeSb$_2$ change under hydrostatic compression. We applied the hydrostatic compression in the pressure range of 0 to 9 GPa, without considering the possibility of the crystal structure phase transition. Figure \ref{fig:pressure-bands}\subref{subfig:Sn-pressure} and \ref{fig:pressure-bands}\subref{subfig:Ge-pressure} show the evolution of bulk energy band structure under hydrostatic pressure, for CdSnSb$_2$ and CdGeSb$_2$, respectively. With increasing pressure, the band structure for both materials undergo similar interesting changes. As discussed earlier, at ambient pressure, the valence bands have major contribution from Sb-$p_{x}$ and Sb-$p_{y}$ orbitals followed by the bands with s-contribution and the conduction bands have major Sb-$p_{z}$ contribution. As the pressure increases, the band gap starts to reduce, while keeping the order of other bands intact and becomes completely zero at the critical pressure of 2.21 and 2.30 GPa for CdSnSb$_2$ and CdGeSb$_2$, respectively. Hence, at low pressures, both the compounds remain TIs with an inverted band order. As compression increases, a topological quantum phase transition occurs and the compounds turn to a Dirac semimetal (DSM) state at the critical pressures. Figure \ref{fig:arpes_and_fermi-Dirac} shows the calculated surface density of states at the critical pressure. Non-trivial edge states merging at the Dirac point can be contrasted to the trivial bulk states.  Interestingly, on further increasing the pressure above this critical point, the energy gap starts to increase along with change in the character of bands. After the critical pressure, the valence and conduction bands have major contribution from Sb-$p_{z}$, and Sb-s orbitals, respectively, which is an inverted order compared to that of the ambient phase. The topological invariant $\mathbb{Z}_{2}$ above the critical point comes out to be zero, thereby confirming the topological quantum phase transition. 

\subsection{Model Hamiltonian} 
Since the orbitals $s$ and $p$ have a major contribution around the $\Gamma$ point at the fermi level, a four-band modified L\"{u}ttinger model\cite{luttinger-original} Hamiltonian is constructed in the basis of $| Sb_{J},j_z \rangle = | Sb_{\frac{3}{2}}, \pm\frac{3}{2} \rangle$ and $| Sb_{\frac{3}{2}}, \pm\frac{1}{2} \rangle$. Mostly, these four states consist of the bonding and antibonding of $s$ and $p$ obitals of Sb atoms.  The Hamiltonian is given by,
\begin{equation}
 H = \frac{\hbar^2}{2m_0} \left( \alpha_1 \mathbf{k}^2 I^0 + \alpha_2 \mathbf{(k.J)}^2 + \alpha_3 \sum \mathbf{k^2J^2}  + \alpha_4 \right)
\label{eq:model}
\end{equation}
where, $I^0$ is a 4 $\times$ 4 identity matrix, and $\mathbf{J}$=(J$_x$, J$_y$, J$_z$) with $J_{i}$ being the $3/2$ angular momentum matrices. The coefficients $\alpha_i$ are L\"{u}ttinger empirical parameters obtained by fitting them with HSE bandstructure. At the $\Gamma$ point, the degeneracy of all four bands is lifted by introducing a band-gap dependent term $\alpha_4$. The first term with quadratic order in momentum in the Eqn. \ref{eq:model} is invariant under spherical symmetry with inversion, representing the isotropic nature of the bands. The second and third terms introduce anisotropy in the bands, lowering the symmetry. 

\begin{figure}[t!]
\centering
\vspace{2mm}
\includegraphics[width=0.8\columnwidth]{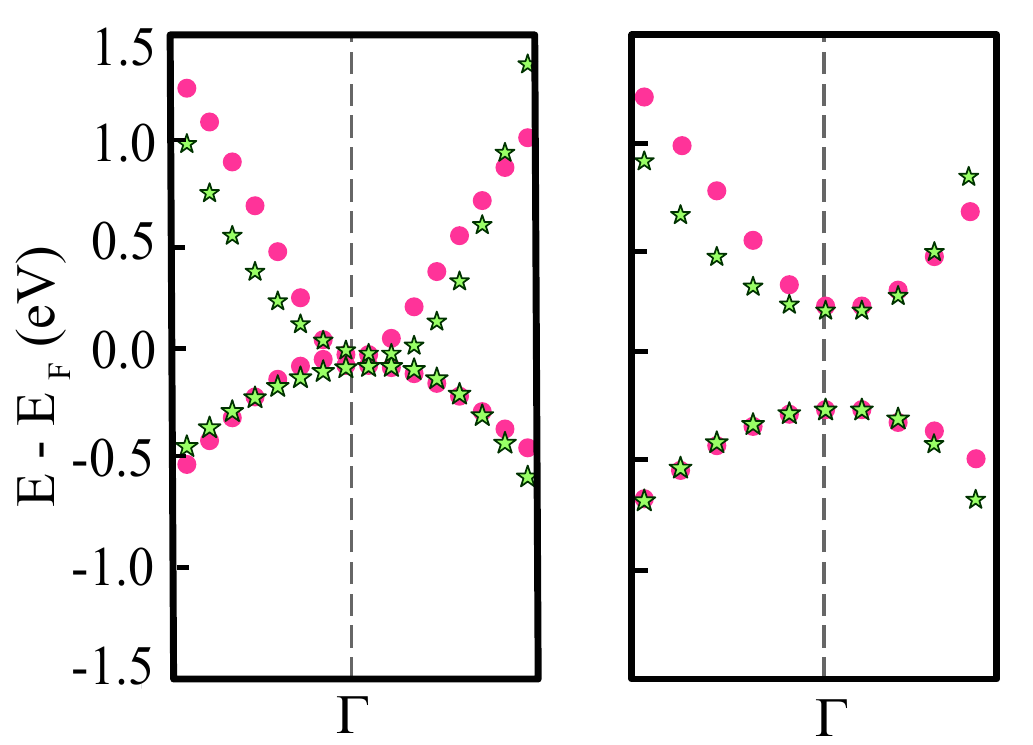}
\vspace{-4mm}
\caption{\label{fig:model-plots} (color online) The L\"{u}ttinger bandstructure for CdSnSb$_2$ fitted to HSE bandstructure for (a) ambient and (b) 8.01 GPa hydrostatic pressure.}
\end{figure}

In the presence of spin-orbit coupling, the above Hamiltonian can be represented by,

\begin{center}
$H = \left( \begin{array}{cccc}
     H_{11}    & H_{12} & H_{13} & H_{14} \\
     H_{12}^*    & H_{22} & H_{23} & H_{24}\\
     H_{13}^*    & H_{23}^* & H_{33} & H_{34} \\
     H_{14}^*    & H_{24}^* & H_{34}^* & H_{44}
\end{array}\right)$
\end{center}

where,
\begin{eqnarray*}
 H_{11} &=&  (\alpha_1 - \frac{3}{4} \alpha_3) (k_x^2+k_y^2)  + (\alpha_1 - \frac{9}{4} \alpha_3) k_z^2 - 3\alpha_2 k_z + \alpha_4 \\
 H_{12} &=& -\sqrt{3}\alpha_2 (k_x - i k_y) \\
 H_{13} &=& -\frac{\sqrt{3}}{2}\alpha_3 (k_x^2 - k_y^2) \\
 H_{14} &=& 0 \\ 
 H_{22} &=& (\alpha_1 - \frac{7}{4} \alpha_3) (k_x^2+k_y^2) +  (\alpha_1 - \frac{1}{4} \alpha_3) k_z^2 -\alpha_2 k_z + \alpha_4 \\
 H_{23} &=& -2 \alpha_2(k_x - i k_y) \\
 H_{24} &=& -\frac{\sqrt{3}}{2} \alpha_3 (k_x^2 - k_y^2) \\
 H_{33} &=& (\alpha_1 - \frac{7}{4} \alpha_3) (k_x^2+k_y^2) +  (\alpha_1 - \frac{1}{4} \alpha_3) k_z^2  + \alpha_2 k_z + \alpha_4 \\
 H_{34} &=& - \sqrt{3} \alpha_2 (k_x - i k_y) \\
 H_{44} &=& (\alpha_1 - \frac{3}{4} \alpha_3) (k_x^2+k_y^2) +  (\alpha_1 - \frac{9}{4} \alpha_3) k_z^2  + 3\alpha_2 k_z + \alpha_4
\end{eqnarray*}
Diagonalization gives two sets of degenerate bands which forms valence and conduction bands.  Figure \ref{fig:model-plots} shows the fitted bands at ambient pressure and 8.01 GPa. The corresponding parameters are listed in Table \ref{tab:model-params}. While the isotropic term remains unchanged, the second and third terms in the Eqn. \ref{eq:model} vary profoundly with pressure, signaling an enhancement of the anisotropy of bands. 

The tunable L\"{u}ttinger parameters can be qualitatively related to the effective masses. Due to the coupling of cross-terms of momenta, the anharmonicity dominates leading to an increase in off-diagonal terms in the effective mass tensor. These tensors for ambient phase, Dirac semimetallic phase, and normal semiconductor phase are given in Appendix \ref{app:emc}. The effective masses in certain directions increase while in some directions decrease, suggesting a built-up of anharmonicity.

The bands for the intermediate pressure values can be obtained from varying the parameters, which are listed in the Table \ref{tab:model-params}. A proper strain engineering enables the control of the topological order. With a suitable choice of parameters, this model can be further extended to study the topological phase transition in a wide range of materials.
\begin{table}[h]
\caption{\label{tab:model-params} L\"{u}ttinger parameters}
\begin{ruledtabular}
\begin{tabular}{lcccc}
    Pressure & $\alpha_1$ (\AA{}$^2$ eV) & $\alpha_2$ (\AA{}$^2$ eV) & $\alpha_3$ (\AA{}$^2$ eV) & $\alpha_4$ (eV)\\
    \hline
    ambient & 46 & -19 & -11 & 0.03 \\
    8.01 GPa & 46 & -40 & 11 & 0.23 
    \end{tabular}
\end{ruledtabular}
\end{table}

\section{Conclusion}

In summary, we have performed first-principles calculations to study the topological phase transitions in chalcopyrite compounds as a function of hydrostatic pressure. These compounds are topological insulators in the native phase with an inverted band order around BZ center. Upon hydrostatic compression, there is a transition from nontrivial TI phase to a Dirac semimetallic state at a critical pressure.  Further increase in pressure drives the materials into a trivial semiconductors along with normal ordering of bands. Different quantum phases are characterized by topological invariants as well as surface states. These quantum phase transitions are further validated by model calculations based on L\"{u}ttinger Hamiltonian, which unravels the critical role played by pressure-induced anisotropy of frontier bands in driving the phase transitions. Such a manoeuvre between various topological phases by hydrostatic pressure can stimulate the search for TQPTs in future experiments.

\acknowledgements

R.S. acknowledges Science and Engineering Research Board, India for a fellowship (PDF/2015/000466). R. J. acknowledges support from DST through INSPIRE fellowship (IF150848). This work is partly supported by the U.S. Army Contract FA5209-16-P-0090, and DST Nanomission. We also acknowledge Materials Research Center and Supercomputer Education and Research Center, Indian Institute of Science for providing the required computational facilities.  

\FloatBarrier 

\appendix
\section{Effective mass tensors}\label{app:emc}
The effective mass tensors of CdSnSb$_2$ at three different phases (i.e. topological insulator, Dirac semimetallic and semiconducting phases respectively) are given below.

\begin{center}
$m_{TI}^{*} =  
 \left( 
\begin{array}{ccc}
   -779.6   & -21.8 & -7.6 \\
      & -571.5 & 4.6\\
      &  & -769.4 \\
\end{array}
\right) $
\vspace{2mm}
$m_{DM}^{*} =  
 \left( 
\begin{array}{ccc}
   -985.5  & -24.8 & -10.1 \\
      & -729.6 & 3.2 \\
      &  & -968.3 
\end{array}
\right) $
\vspace{2mm}
$m_{SM}^{*} =  
 \left( 
\begin{array}{ccc}
  -1049.4   & -26.2 & -11.1 \\
      & -775.3 & 2.8 \\
      &  & -1029.5 
\end{array}
\right) $

\end{center}
\vspace{3cm}

\begin{thebibliography}{54}%
\makeatletter
\providecommand \@ifxundefined [1]{%
 \@ifx{#1\undefined}
}%
\providecommand \@ifnum [1]{%
 \ifnum #1\expandafter \@firstoftwo
 \else \expandafter \@secondoftwo
 \fi
}%
\providecommand \@ifx [1]{%
 \ifx #1\expandafter \@firstoftwo
 \else \expandafter \@secondoftwo
 \fi
}%
\providecommand \natexlab [1]{#1}%
\providecommand \enquote  [1]{``#1''}%
\providecommand \bibnamefont  [1]{#1}%
\providecommand \bibfnamefont [1]{#1}%
\providecommand \citenamefont [1]{#1}%
\providecommand \href@noop [0]{\@secondoftwo}%
\providecommand \href [0]{\begingroup \@sanitize@url \@href}%
\providecommand \@href[1]{\@@startlink{#1}\@@href}%
\providecommand \@@href[1]{\endgroup#1\@@endlink}%
\providecommand \@sanitize@url [0]{\catcode `\\12\catcode `\$12\catcode
  `\&12\catcode `\#12\catcode `\^12\catcode `\_12\catcode `\%12\relax}%
\providecommand \@@startlink[1]{}%
\providecommand \@@endlink[0]{}%
\providecommand \url  [0]{\begingroup\@sanitize@url \@url }%
\providecommand \@url [1]{\endgroup\@href {#1}{\urlprefix }}%
\providecommand \urlprefix  [0]{URL }%
\providecommand \Eprint [0]{\href }%
\providecommand \doibase [0]{http://dx.doi.org/}%
\providecommand \selectlanguage [0]{\@gobble}%
\providecommand \bibinfo  [0]{\@secondoftwo}%
\providecommand \bibfield  [0]{\@secondoftwo}%
\providecommand \translation [1]{[#1]}%
\providecommand \BibitemOpen [0]{}%
\providecommand \bibitemStop [0]{}%
\providecommand \bibitemNoStop [0]{.\EOS\space}%
\providecommand \EOS [0]{\spacefactor3000\relax}%
\providecommand \BibitemShut  [1]{\csname bibitem#1\endcsname}%
\let\auto@bib@innerbib\@empty
\bibitem [{\citenamefont {Kane}\ and\ \citenamefont
  {Mele}(2005)}]{2005Prl-Kane-Z2}%
  \BibitemOpen
  \bibfield  {author} {\bibinfo {author} {\bibfnamefont {C.~L.}\ \bibnamefont
  {Kane}}\ and\ \bibinfo {author} {\bibfnamefont {E.~J.}\ \bibnamefont
  {Mele}},\ }\href {\doibase 10.1103/PhysRevLett.95.146802} {\bibfield
  {journal} {\bibinfo  {journal} {Phys. Rev. Lett.}\ }\textbf {\bibinfo
  {volume} {95}},\ \bibinfo {pages} {146802} (\bibinfo {year}
  {2005})}\BibitemShut {NoStop}%
\bibitem [{\citenamefont {Fu}\ \emph {et~al.}(2007)\citenamefont {Fu},
  \citenamefont {Kane},\ and\ \citenamefont {Mele}}]{2007Prl-Fu-TI3D}%
  \BibitemOpen
  \bibfield  {author} {\bibinfo {author} {\bibfnamefont {L.}~\bibnamefont
  {Fu}}, \bibinfo {author} {\bibfnamefont {C.~L.}\ \bibnamefont {Kane}}, \ and\
  \bibinfo {author} {\bibfnamefont {E.~J.}\ \bibnamefont {Mele}},\ }\href
  {\doibase 10.1103/PhysRevLett.98.106803} {\bibfield  {journal} {\bibinfo
  {journal} {Phys. Rev. Lett.}\ }\textbf {\bibinfo {volume} {98}},\ \bibinfo
  {pages} {106803} (\bibinfo {year} {2007})}\BibitemShut {NoStop}%
\bibitem [{\citenamefont {Hasan}\ and\ \citenamefont
  {Kane}(2010)}]{2010ReviewMP-Hasan-colloquium}%
  \BibitemOpen
  \bibfield  {author} {\bibinfo {author} {\bibfnamefont {M.~Z.}\ \bibnamefont
  {Hasan}}\ and\ \bibinfo {author} {\bibfnamefont {C.~L.}\ \bibnamefont
  {Kane}},\ }\href {\doibase 10.1103/RevModPhys.82.3045} {\bibfield  {journal}
  {\bibinfo  {journal} {Rev. Mod. Phys.}\ }\textbf {\bibinfo {volume} {82}},\
  \bibinfo {pages} {3045} (\bibinfo {year} {2010})}\BibitemShut {NoStop}%
\bibitem [{\citenamefont {Bernevig}\ \emph {et~al.}(2006)\citenamefont
  {Bernevig}, \citenamefont {Hughes},\ and\ \citenamefont
  {Zhang}}]{2006science-Bernevig-HgTe}%
  \BibitemOpen
  \bibfield  {author} {\bibinfo {author} {\bibfnamefont {B.~A.}\ \bibnamefont
  {Bernevig}}, \bibinfo {author} {\bibfnamefont {T.~L.}\ \bibnamefont
  {Hughes}}, \ and\ \bibinfo {author} {\bibfnamefont {S.-C.}\ \bibnamefont
  {Zhang}},\ }\href {\doibase 10.1126/science.1133734} {\bibfield  {journal}
  {\bibinfo  {journal} {Science}\ }\textbf {\bibinfo {volume} {314}},\ \bibinfo
  {pages} {1757} (\bibinfo {year} {2006})}\BibitemShut {NoStop}%
\bibitem [{\citenamefont {Hsieh}\ \emph {et~al.}(2008)\citenamefont {Hsieh},
  \citenamefont {Qian}, \citenamefont {Wray}, \citenamefont {Xia},
  \citenamefont {Hor}, \citenamefont {Cava},\ and\ \citenamefont
  {Hasan}}]{2008nature-Hasan-DSM}%
  \BibitemOpen
  \bibfield  {author} {\bibinfo {author} {\bibfnamefont {D.}~\bibnamefont
  {Hsieh}}, \bibinfo {author} {\bibfnamefont {D.}~\bibnamefont {Qian}},
  \bibinfo {author} {\bibfnamefont {L.}~\bibnamefont {Wray}}, \bibinfo {author}
  {\bibfnamefont {Y.}~\bibnamefont {Xia}}, \bibinfo {author} {\bibfnamefont
  {Y.~S.}\ \bibnamefont {Hor}}, \bibinfo {author} {\bibfnamefont {R.~J.}\
  \bibnamefont {Cava}}, \ and\ \bibinfo {author} {\bibfnamefont {M.~Z.}\
  \bibnamefont {Hasan}},\ }\href {\doibase 10.1038/nature06843} {\bibfield
  {journal} {\bibinfo  {journal} {Nature}\ }\textbf {\bibinfo {volume} {452}},\
  \bibinfo {pages} {970} (\bibinfo {year} {2008})}\BibitemShut {NoStop}%
\bibitem [{\citenamefont {Teo}\ \emph {et~al.}(2008)\citenamefont {Teo},
  \citenamefont {Fu},\ and\ \citenamefont {Kane}}]{2008PRB-Kane-BiSb}%
  \BibitemOpen
  \bibfield  {author} {\bibinfo {author} {\bibfnamefont {J.~C.~Y.}\
  \bibnamefont {Teo}}, \bibinfo {author} {\bibfnamefont {L.}~\bibnamefont
  {Fu}}, \ and\ \bibinfo {author} {\bibfnamefont {C.~L.}\ \bibnamefont
  {Kane}},\ }\href {https://doi.org/10.1103/physrevb.78.045426} {\bibfield
  {journal} {\bibinfo  {journal} {Phys. Rev. B}\ }\textbf {\bibinfo {volume}
  {78}} (\bibinfo {year} {2008})}\BibitemShut {NoStop}%
\bibitem [{\citenamefont {Zhang}\ \emph {et~al.}(2009)\citenamefont {Zhang},
  \citenamefont {Liu}, \citenamefont {Qi}, \citenamefont {Dai}, \citenamefont
  {Fang},\ and\ \citenamefont {Zhang}}]{2009naturephysics-Zhang-TIs}%
  \BibitemOpen
  \bibfield  {author} {\bibinfo {author} {\bibfnamefont {H.}~\bibnamefont
  {Zhang}}, \bibinfo {author} {\bibfnamefont {C.-X.}\ \bibnamefont {Liu}},
  \bibinfo {author} {\bibfnamefont {X.-L.}\ \bibnamefont {Qi}}, \bibinfo
  {author} {\bibfnamefont {X.}~\bibnamefont {Dai}}, \bibinfo {author}
  {\bibfnamefont {Z.}~\bibnamefont {Fang}}, \ and\ \bibinfo {author}
  {\bibfnamefont {S.-C.}\ \bibnamefont {Zhang}},\ }\href {\doibase
  10.1038/nphys1270} {\bibfield  {journal} {\bibinfo  {journal} {Nat. Phys.}\
  }\textbf {\bibinfo {volume} {5}},\ \bibinfo {pages} {438} (\bibinfo {year}
  {2009})}\BibitemShut {NoStop}%
\bibitem [{\citenamefont {Chen}\ \emph {et~al.}(2009)\citenamefont {Chen},
  \citenamefont {Analytis}, \citenamefont {Chu}, \citenamefont {Liu},
  \citenamefont {Mo}, \citenamefont {Qi}, \citenamefont {Zhang}, \citenamefont
  {Lu}, \citenamefont {Dai}, \citenamefont {Fang}, \citenamefont {Zhang},
  \citenamefont {Fisher}, \citenamefont {Hussain},\ and\ \citenamefont
  {Shen}}]{2009science-Chen-Bi2Te3}%
  \BibitemOpen
  \bibfield  {author} {\bibinfo {author} {\bibfnamefont {Y.~L.}\ \bibnamefont
  {Chen}}, \bibinfo {author} {\bibfnamefont {J.~G.}\ \bibnamefont {Analytis}},
  \bibinfo {author} {\bibfnamefont {J.-H.}\ \bibnamefont {Chu}}, \bibinfo
  {author} {\bibfnamefont {Z.~K.}\ \bibnamefont {Liu}}, \bibinfo {author}
  {\bibfnamefont {S.-K.}\ \bibnamefont {Mo}}, \bibinfo {author} {\bibfnamefont
  {X.~L.}\ \bibnamefont {Qi}}, \bibinfo {author} {\bibfnamefont {H.~J.}\
  \bibnamefont {Zhang}}, \bibinfo {author} {\bibfnamefont {D.~H.}\ \bibnamefont
  {Lu}}, \bibinfo {author} {\bibfnamefont {X.}~\bibnamefont {Dai}}, \bibinfo
  {author} {\bibfnamefont {Z.}~\bibnamefont {Fang}}, \bibinfo {author}
  {\bibfnamefont {S.~C.}\ \bibnamefont {Zhang}}, \bibinfo {author}
  {\bibfnamefont {I.~R.}\ \bibnamefont {Fisher}}, \bibinfo {author}
  {\bibfnamefont {Z.}~\bibnamefont {Hussain}}, \ and\ \bibinfo {author}
  {\bibfnamefont {Z.-X.}\ \bibnamefont {Shen}},\ }\href {\doibase
  10.1126/science.1173034} {\bibfield  {journal} {\bibinfo  {journal}
  {Science}\ }\textbf {\bibinfo {volume} {325}},\ \bibinfo {pages} {178}
  (\bibinfo {year} {2009})}\BibitemShut {NoStop}%
\bibitem [{\citenamefont {Zhong}\ \emph {et~al.}(2014)\citenamefont {Zhong},
  \citenamefont {Schneeloch}, \citenamefont {Liu}, \citenamefont {Camino},
  \citenamefont {Tranquada},\ and\ \citenamefont
  {Gu}}]{2014PRB-Zhong-superconductorPbSnTe}%
  \BibitemOpen
  \bibfield  {author} {\bibinfo {author} {\bibfnamefont {R.~D.}\ \bibnamefont
  {Zhong}}, \bibinfo {author} {\bibfnamefont {J.~A.}\ \bibnamefont
  {Schneeloch}}, \bibinfo {author} {\bibfnamefont {T.~S.}\ \bibnamefont {Liu}},
  \bibinfo {author} {\bibfnamefont {F.~E.}\ \bibnamefont {Camino}}, \bibinfo
  {author} {\bibfnamefont {J.~M.}\ \bibnamefont {Tranquada}}, \ and\ \bibinfo
  {author} {\bibfnamefont {G.~D.}\ \bibnamefont {Gu}},\ }\href {\doibase
  10.1103/PhysRevB.90.020505} {\bibfield  {journal} {\bibinfo  {journal} {Phys.
  Rev. B}\ }\textbf {\bibinfo {volume} {90}},\ \bibinfo {pages} {020505}
  (\bibinfo {year} {2014})}\BibitemShut {NoStop}%
\bibitem [{\citenamefont {Qi}\ and\ \citenamefont
  {Zhang}(2011)}]{2011ReviewMP-Qi-TIs}%
  \BibitemOpen
  \bibfield  {author} {\bibinfo {author} {\bibfnamefont {X.-L.}\ \bibnamefont
  {Qi}}\ and\ \bibinfo {author} {\bibfnamefont {S.-C.}\ \bibnamefont {Zhang}},\
  }\href {\doibase 10.1103/revmodphys.83.1057} {\bibfield  {journal} {\bibinfo
  {journal} {Reviews of Modern Physics}\ }\textbf {\bibinfo {volume} {83}},\
  \bibinfo {pages} {1057} (\bibinfo {year} {2011})}\BibitemShut {NoStop}%
\bibitem [{\citenamefont {Liu}\ \emph {et~al.}(2014)\citenamefont {Liu},
  \citenamefont {Zhou}, \citenamefont {Zhang}, \citenamefont {Wang},
  \citenamefont {Weng}, \citenamefont {Prabhakaran}, \citenamefont {Mo},
  \citenamefont {Shen}, \citenamefont {Fang}, \citenamefont {Dai},
  \citenamefont {Hussain},\ and\ \citenamefont
  {Chen}}]{2014science-Liu-diracsemimetalNa3Bi}%
  \BibitemOpen
  \bibfield  {author} {\bibinfo {author} {\bibfnamefont {Z.~K.}\ \bibnamefont
  {Liu}}, \bibinfo {author} {\bibfnamefont {B.}~\bibnamefont {Zhou}}, \bibinfo
  {author} {\bibfnamefont {Y.}~\bibnamefont {Zhang}}, \bibinfo {author}
  {\bibfnamefont {Z.~J.}\ \bibnamefont {Wang}}, \bibinfo {author}
  {\bibfnamefont {H.~M.}\ \bibnamefont {Weng}}, \bibinfo {author}
  {\bibfnamefont {D.}~\bibnamefont {Prabhakaran}}, \bibinfo {author}
  {\bibfnamefont {S.-K.}\ \bibnamefont {Mo}}, \bibinfo {author} {\bibfnamefont
  {Z.~X.}\ \bibnamefont {Shen}}, \bibinfo {author} {\bibfnamefont
  {Z.}~\bibnamefont {Fang}}, \bibinfo {author} {\bibfnamefont {X.}~\bibnamefont
  {Dai}}, \bibinfo {author} {\bibfnamefont {Z.}~\bibnamefont {Hussain}}, \ and\
  \bibinfo {author} {\bibfnamefont {Y.~L.}\ \bibnamefont {Chen}},\ }\href
  {\doibase 10.1126/science.1245085} {\bibfield  {journal} {\bibinfo  {journal}
  {Science}\ }\textbf {\bibinfo {volume} {343}},\ \bibinfo {pages} {864}
  (\bibinfo {year} {2014})}\BibitemShut {NoStop}%
\bibitem [{\citenamefont {Neupane}\ \emph {et~al.}(2014)\citenamefont
  {Neupane}, \citenamefont {Xu}, \citenamefont {Sankar}, \citenamefont
  {Alidoust}, \citenamefont {Bian}, \citenamefont {Liu}, \citenamefont
  {Belopolski}, \citenamefont {Chang}, \citenamefont {Jeng}, \citenamefont
  {Lin}, \citenamefont {Bansil}, \citenamefont {Chou},\ and\ \citenamefont
  {Hasan}}]{2014naturecommun-Neupane-diracsemimetalCd3As2}%
  \BibitemOpen
  \bibfield  {author} {\bibinfo {author} {\bibfnamefont {M.}~\bibnamefont
  {Neupane}}, \bibinfo {author} {\bibfnamefont {S.-Y.}\ \bibnamefont {Xu}},
  \bibinfo {author} {\bibfnamefont {R.}~\bibnamefont {Sankar}}, \bibinfo
  {author} {\bibfnamefont {N.}~\bibnamefont {Alidoust}}, \bibinfo {author}
  {\bibfnamefont {G.}~\bibnamefont {Bian}}, \bibinfo {author} {\bibfnamefont
  {C.}~\bibnamefont {Liu}}, \bibinfo {author} {\bibfnamefont {I.}~\bibnamefont
  {Belopolski}}, \bibinfo {author} {\bibfnamefont {T.-R.}\ \bibnamefont
  {Chang}}, \bibinfo {author} {\bibfnamefont {H.-T.}\ \bibnamefont {Jeng}},
  \bibinfo {author} {\bibfnamefont {H.}~\bibnamefont {Lin}}, \bibinfo {author}
  {\bibfnamefont {A.}~\bibnamefont {Bansil}}, \bibinfo {author} {\bibfnamefont
  {F.}~\bibnamefont {Chou}}, \ and\ \bibinfo {author} {\bibfnamefont {M.~Z.}\
  \bibnamefont {Hasan}},\ }\href {http://dx.doi.org/10.1038/ncomms4786}
  {\bibfield  {journal} {\bibinfo  {journal} {Nat. Commun.}\ }\textbf {\bibinfo
  {volume} {5}} (\bibinfo {year} {2014})}\BibitemShut {NoStop}%
\bibitem [{\citenamefont {Wang}\ \emph {et~al.}(2012)\citenamefont {Wang},
  \citenamefont {Sun}, \citenamefont {Chen}, \citenamefont {Franchini},
  \citenamefont {Xu}, \citenamefont {Weng}, \citenamefont {Dai},\ and\
  \citenamefont {Fang}}]{2012PRB-Wang-A3Bi-dirac}%
  \BibitemOpen
  \bibfield  {author} {\bibinfo {author} {\bibfnamefont {Z.}~\bibnamefont
  {Wang}}, \bibinfo {author} {\bibfnamefont {Y.}~\bibnamefont {Sun}}, \bibinfo
  {author} {\bibfnamefont {X.-Q.}\ \bibnamefont {Chen}}, \bibinfo {author}
  {\bibfnamefont {C.}~\bibnamefont {Franchini}}, \bibinfo {author}
  {\bibfnamefont {G.}~\bibnamefont {Xu}}, \bibinfo {author} {\bibfnamefont
  {H.}~\bibnamefont {Weng}}, \bibinfo {author} {\bibfnamefont {X.}~\bibnamefont
  {Dai}}, \ and\ \bibinfo {author} {\bibfnamefont {Z.}~\bibnamefont {Fang}},\
  }\href {\doibase 10.1103/PhysRevB.85.195320} {\bibfield  {journal} {\bibinfo
  {journal} {Phys. Rev. B}\ }\textbf {\bibinfo {volume} {85}},\ \bibinfo
  {pages} {195320} (\bibinfo {year} {2012})}\BibitemShut {NoStop}%
\bibitem [{\citenamefont {Yang}\ and\ \citenamefont
  {Nagaosa}(2014)}]{2014Naturecommun-Yang-diracsemimetal}%
  \BibitemOpen
  \bibfield  {author} {\bibinfo {author} {\bibfnamefont {B.-J.}\ \bibnamefont
  {Yang}}\ and\ \bibinfo {author} {\bibfnamefont {N.}~\bibnamefont {Nagaosa}},\
  }\href {\doibase 10.1038/ncomms5898} {\bibfield  {journal} {\bibinfo
  {journal} {Nat. Commun.}\ }\textbf {\bibinfo {volume} {5}},\ \bibinfo {pages}
  {4898} (\bibinfo {year} {2014})}\BibitemShut {NoStop}%
\bibitem [{\citenamefont {Gupta}\ \emph {et~al.}(2017)\citenamefont {Gupta},
  \citenamefont {Juneja}, \citenamefont {Shinde},\ and\ \citenamefont
  {Singh}}]{thsim-sunny}%
  \BibitemOpen
  \bibfield  {author} {\bibinfo {author} {\bibfnamefont {S.}~\bibnamefont
  {Gupta}}, \bibinfo {author} {\bibfnamefont {R.}~\bibnamefont {Juneja}},
  \bibinfo {author} {\bibfnamefont {R.}~\bibnamefont {Shinde}}, \ and\ \bibinfo
  {author} {\bibfnamefont {A.~K.}\ \bibnamefont {Singh}},\ }\href {\doibase
  10.1063/1.4984262} {\bibfield  {journal} {\bibinfo  {journal} {J. Appl.
  Phys.}\ }\textbf {\bibinfo {volume} {121}},\ \bibinfo {pages} {214901}
  (\bibinfo {year} {2017})}\BibitemShut {NoStop}%
\bibitem [{\citenamefont {Barik}\ \emph {et~al.}(2018)\citenamefont {Barik},
  \citenamefont {Shinde},\ and\ \citenamefont {Singh}}]{thsim-ranjan}%
  \BibitemOpen
  \bibfield  {author} {\bibinfo {author} {\bibfnamefont {R.~K.}\ \bibnamefont
  {Barik}}, \bibinfo {author} {\bibfnamefont {R.}~\bibnamefont {Shinde}}, \
  and\ \bibinfo {author} {\bibfnamefont {A.~K.}\ \bibnamefont {Singh}},\
  }\href@noop {} {\enquote {\bibinfo {title} {Multiple triple-point fermions in
  heusler compounds},}\ } (\bibinfo {year} {2018}),\ \Eprint
  {http://arxiv.org/abs/arXiv:1801.05699} {arXiv:1801.05699} \BibitemShut
  {NoStop}%
\bibitem [{\citenamefont {Xu}\ \emph {et~al.}(2011)\citenamefont {Xu},
  \citenamefont {Xia}, \citenamefont {Wray}, \citenamefont {Jia}, \citenamefont
  {Meier}, \citenamefont {Dil}, \citenamefont {Osterwalder}, \citenamefont
  {Slomski}, \citenamefont {Bansil}, \citenamefont {Lin}, \citenamefont
  {Cava},\ and\ \citenamefont {Hasan}}]{2011-Science-hasan-texture-inversion}%
  \BibitemOpen
  \bibfield  {author} {\bibinfo {author} {\bibfnamefont {S.-Y.}\ \bibnamefont
  {Xu}}, \bibinfo {author} {\bibfnamefont {Y.}~\bibnamefont {Xia}}, \bibinfo
  {author} {\bibfnamefont {L.~A.}\ \bibnamefont {Wray}}, \bibinfo {author}
  {\bibfnamefont {S.}~\bibnamefont {Jia}}, \bibinfo {author} {\bibfnamefont
  {F.}~\bibnamefont {Meier}}, \bibinfo {author} {\bibfnamefont {J.~H.}\
  \bibnamefont {Dil}}, \bibinfo {author} {\bibfnamefont {J.}~\bibnamefont
  {Osterwalder}}, \bibinfo {author} {\bibfnamefont {B.}~\bibnamefont
  {Slomski}}, \bibinfo {author} {\bibfnamefont {A.}~\bibnamefont {Bansil}},
  \bibinfo {author} {\bibfnamefont {H.}~\bibnamefont {Lin}}, \bibinfo {author}
  {\bibfnamefont {R.~J.}\ \bibnamefont {Cava}}, \ and\ \bibinfo {author}
  {\bibfnamefont {M.~Z.}\ \bibnamefont {Hasan}},\ }\href {\doibase
  10.1126/science.1201607} {\bibfield  {journal} {\bibinfo  {journal}
  {Science}\ }\textbf {\bibinfo {volume} {332}},\ \bibinfo {pages} {560}
  (\bibinfo {year} {2011})}\BibitemShut {NoStop}%
\bibitem [{\citenamefont {Brahlek}\ \emph {et~al.}(2012)\citenamefont
  {Brahlek}, \citenamefont {Bansal}, \citenamefont {Koirala}, \citenamefont
  {Xu}, \citenamefont {Neupane}, \citenamefont {Liu}, \citenamefont {Hasan},\
  and\ \citenamefont {Oh}}]{2012Prl-Brahlek-BiInSe-TI-transition}%
  \BibitemOpen
  \bibfield  {author} {\bibinfo {author} {\bibfnamefont {M.}~\bibnamefont
  {Brahlek}}, \bibinfo {author} {\bibfnamefont {N.}~\bibnamefont {Bansal}},
  \bibinfo {author} {\bibfnamefont {N.}~\bibnamefont {Koirala}}, \bibinfo
  {author} {\bibfnamefont {S.-Y.}\ \bibnamefont {Xu}}, \bibinfo {author}
  {\bibfnamefont {M.}~\bibnamefont {Neupane}}, \bibinfo {author} {\bibfnamefont
  {C.}~\bibnamefont {Liu}}, \bibinfo {author} {\bibfnamefont {M.~Z.}\
  \bibnamefont {Hasan}}, \ and\ \bibinfo {author} {\bibfnamefont
  {S.}~\bibnamefont {Oh}},\ }\href {\doibase 10.1103/PhysRevLett.109.186403}
  {\bibfield  {journal} {\bibinfo  {journal} {Phys. Rev. Lett.}\ }\textbf
  {\bibinfo {volume} {109}},\ \bibinfo {pages} {186403} (\bibinfo {year}
  {2012})}\BibitemShut {NoStop}%
\bibitem [{\citenamefont {Sato}\ \emph {et~al.}(2011)\citenamefont {Sato},
  \citenamefont {Segawa}, \citenamefont {Kosaka}, \citenamefont {Souma},
  \citenamefont {Nakayama}, \citenamefont {Eto}, \citenamefont {Minami},
  \citenamefont {Ando},\ and\ \citenamefont
  {Takahashi}}]{2011-NatPhy-Sato-QPT}%
  \BibitemOpen
  \bibfield  {author} {\bibinfo {author} {\bibfnamefont {T.}~\bibnamefont
  {Sato}}, \bibinfo {author} {\bibfnamefont {K.}~\bibnamefont {Segawa}},
  \bibinfo {author} {\bibfnamefont {K.}~\bibnamefont {Kosaka}}, \bibinfo
  {author} {\bibfnamefont {S.}~\bibnamefont {Souma}}, \bibinfo {author}
  {\bibfnamefont {K.}~\bibnamefont {Nakayama}}, \bibinfo {author}
  {\bibfnamefont {K.}~\bibnamefont {Eto}}, \bibinfo {author} {\bibfnamefont
  {T.}~\bibnamefont {Minami}}, \bibinfo {author} {\bibfnamefont
  {Y.}~\bibnamefont {Ando}}, \ and\ \bibinfo {author} {\bibfnamefont
  {T.}~\bibnamefont {Takahashi}},\ }\href {\doibase 10.1038/nphys2058}
  {\bibfield  {journal} {\bibinfo  {journal} {Nat. Phys.}\ }\textbf {\bibinfo
  {volume} {7}},\ \bibinfo {pages} {840} (\bibinfo {year} {2011})}\BibitemShut
  {NoStop}%
\bibitem [{\citenamefont {Wu}\ \emph {et~al.}(2013)\citenamefont {Wu},
  \citenamefont {Brahlek}, \citenamefont {Aguilar}, \citenamefont {Stier},
  \citenamefont {Morris}, \citenamefont {Lubashevsky}, \citenamefont {Bilbro},
  \citenamefont {Bansal}, \citenamefont {Oh},\ and\ \citenamefont
  {Armitage}}]{2013-NatPhy-Wu-TPT}%
  \BibitemOpen
  \bibfield  {author} {\bibinfo {author} {\bibfnamefont {L.}~\bibnamefont
  {Wu}}, \bibinfo {author} {\bibfnamefont {M.}~\bibnamefont {Brahlek}},
  \bibinfo {author} {\bibfnamefont {R.~V.}\ \bibnamefont {Aguilar}}, \bibinfo
  {author} {\bibfnamefont {A.~V.}\ \bibnamefont {Stier}}, \bibinfo {author}
  {\bibfnamefont {C.~M.}\ \bibnamefont {Morris}}, \bibinfo {author}
  {\bibfnamefont {Y.}~\bibnamefont {Lubashevsky}}, \bibinfo {author}
  {\bibfnamefont {L.~S.}\ \bibnamefont {Bilbro}}, \bibinfo {author}
  {\bibfnamefont {N.}~\bibnamefont {Bansal}}, \bibinfo {author} {\bibfnamefont
  {S.}~\bibnamefont {Oh}}, \ and\ \bibinfo {author} {\bibfnamefont {N.~P.}\
  \bibnamefont {Armitage}},\ }\href {\doibase 10.1038/nphys2647} {\bibfield
  {journal} {\bibinfo  {journal} {Nat. Phys.}\ }\textbf {\bibinfo {volume}
  {9}},\ \bibinfo {pages} {410} (\bibinfo {year} {2013})}\BibitemShut {NoStop}%
\bibitem [{\citenamefont {Yan}\ \emph {et~al.}(2014)\citenamefont {Yan},
  \citenamefont {Liu}, \citenamefont {Zang}, \citenamefont {Wang},
  \citenamefont {Wang}, \citenamefont {Wang}, \citenamefont {Zhang},
  \citenamefont {Wang}, \citenamefont {Ma}, \citenamefont {Ji}, \citenamefont
  {He}, \citenamefont {Fu}, \citenamefont {Duan}, \citenamefont {Xue},\ and\
  \citenamefont {Chen}}]{2014-PRL-Yan-doping-transition}%
  \BibitemOpen
  \bibfield  {author} {\bibinfo {author} {\bibfnamefont {C.}~\bibnamefont
  {Yan}}, \bibinfo {author} {\bibfnamefont {J.}~\bibnamefont {Liu}}, \bibinfo
  {author} {\bibfnamefont {Y.}~\bibnamefont {Zang}}, \bibinfo {author}
  {\bibfnamefont {J.}~\bibnamefont {Wang}}, \bibinfo {author} {\bibfnamefont
  {Z.}~\bibnamefont {Wang}}, \bibinfo {author} {\bibfnamefont {P.}~\bibnamefont
  {Wang}}, \bibinfo {author} {\bibfnamefont {Z.-D.}\ \bibnamefont {Zhang}},
  \bibinfo {author} {\bibfnamefont {L.}~\bibnamefont {Wang}}, \bibinfo {author}
  {\bibfnamefont {X.}~\bibnamefont {Ma}}, \bibinfo {author} {\bibfnamefont
  {S.}~\bibnamefont {Ji}}, \bibinfo {author} {\bibfnamefont {K.}~\bibnamefont
  {He}}, \bibinfo {author} {\bibfnamefont {L.}~\bibnamefont {Fu}}, \bibinfo
  {author} {\bibfnamefont {W.}~\bibnamefont {Duan}}, \bibinfo {author}
  {\bibfnamefont {Q.-K.}\ \bibnamefont {Xue}}, \ and\ \bibinfo {author}
  {\bibfnamefont {X.}~\bibnamefont {Chen}},\ }\href
  {https://doi.org/10.1103/physrevlett.112.186801} {\bibfield  {journal}
  {\bibinfo  {journal} {Phys. Rev. Lett.}\ }\textbf {\bibinfo {volume} {112}}
  (\bibinfo {year} {2014})}\BibitemShut {NoStop}%
\bibitem [{\citenamefont {Novak}\ \emph {et~al.}(2015)\citenamefont {Novak},
  \citenamefont {Sasaki}, \citenamefont {Segawa},\ and\ \citenamefont
  {Ando}}]{2015PRB-Novak-TlBiSSe}%
  \BibitemOpen
  \bibfield  {author} {\bibinfo {author} {\bibfnamefont {M.}~\bibnamefont
  {Novak}}, \bibinfo {author} {\bibfnamefont {S.}~\bibnamefont {Sasaki}},
  \bibinfo {author} {\bibfnamefont {K.}~\bibnamefont {Segawa}}, \ and\ \bibinfo
  {author} {\bibfnamefont {Y.}~\bibnamefont {Ando}},\ }\href {\doibase
  10.1103/PhysRevB.91.041203} {\bibfield  {journal} {\bibinfo  {journal} {Phys.
  Rev. B}\ }\textbf {\bibinfo {volume} {91}},\ \bibinfo {pages} {041203}
  (\bibinfo {year} {2015})}\BibitemShut {NoStop}%
\bibitem [{\citenamefont {Singh}\ \emph {et~al.}(2012)\citenamefont {Singh},
  \citenamefont {Sharma}, \citenamefont {Lin}, \citenamefont {Hasan},
  \citenamefont {Prasad},\ and\ \citenamefont
  {Bansil}}]{2012PRB-Singh-TlBiSe2}%
  \BibitemOpen
  \bibfield  {author} {\bibinfo {author} {\bibfnamefont {B.}~\bibnamefont
  {Singh}}, \bibinfo {author} {\bibfnamefont {A.}~\bibnamefont {Sharma}},
  \bibinfo {author} {\bibfnamefont {H.}~\bibnamefont {Lin}}, \bibinfo {author}
  {\bibfnamefont {M.~Z.}\ \bibnamefont {Hasan}}, \bibinfo {author}
  {\bibfnamefont {R.}~\bibnamefont {Prasad}}, \ and\ \bibinfo {author}
  {\bibfnamefont {A.}~\bibnamefont {Bansil}},\ }\href {\doibase
  10.1103/PhysRevB.86.115208} {\bibfield  {journal} {\bibinfo  {journal} {Phys.
  Rev. B}\ }\textbf {\bibinfo {volume} {86}},\ \bibinfo {pages} {115208}
  (\bibinfo {year} {2012})}\BibitemShut {NoStop}%
\bibitem [{\citenamefont {Orlita}\ \emph {et~al.}(2014)\citenamefont {Orlita},
  \citenamefont {Basko}, \citenamefont {Zholudev}, \citenamefont {Teppe},
  \citenamefont {Knap}, \citenamefont {Gavrilenko}, \citenamefont {Mikhailov},
  \citenamefont {Dvoretskii}, \citenamefont {Neugebauer}, \citenamefont
  {Faugeras}, \citenamefont {Barra}, \citenamefont {Martinez},\ and\
  \citenamefont {Potemski}}]{2014Naturephysics-Orlita-3Dkane-fermions}%
  \BibitemOpen
  \bibfield  {author} {\bibinfo {author} {\bibfnamefont {M.}~\bibnamefont
  {Orlita}}, \bibinfo {author} {\bibfnamefont {D.~M.}\ \bibnamefont {Basko}},
  \bibinfo {author} {\bibfnamefont {M.~S.}\ \bibnamefont {Zholudev}}, \bibinfo
  {author} {\bibfnamefont {F.}~\bibnamefont {Teppe}}, \bibinfo {author}
  {\bibfnamefont {W.}~\bibnamefont {Knap}}, \bibinfo {author} {\bibfnamefont
  {V.~I.}\ \bibnamefont {Gavrilenko}}, \bibinfo {author} {\bibfnamefont
  {N.~N.}\ \bibnamefont {Mikhailov}}, \bibinfo {author} {\bibfnamefont {S.~A.}\
  \bibnamefont {Dvoretskii}}, \bibinfo {author} {\bibfnamefont
  {P.}~\bibnamefont {Neugebauer}}, \bibinfo {author} {\bibfnamefont
  {C.}~\bibnamefont {Faugeras}}, \bibinfo {author} {\bibfnamefont {A.-L.}\
  \bibnamefont {Barra}}, \bibinfo {author} {\bibfnamefont {G.}~\bibnamefont
  {Martinez}}, \ and\ \bibinfo {author} {\bibfnamefont {M.}~\bibnamefont
  {Potemski}},\ }\href {\doibase 10.1038/nphys2857} {\bibfield  {journal}
  {\bibinfo  {journal} {Nat. Phys.}\ }\textbf {\bibinfo {volume} {10}},\
  \bibinfo {pages} {233} (\bibinfo {year} {2014})}\BibitemShut {NoStop}%
\bibitem [{\citenamefont {Agapito}\ \emph {et~al.}(2013)\citenamefont
  {Agapito}, \citenamefont {Kioussis}, \citenamefont {Goddard},\ and\
  \citenamefont {Ong}}]{2013-PRL-goddard-tellurium}%
  \BibitemOpen
  \bibfield  {author} {\bibinfo {author} {\bibfnamefont {L.~A.}\ \bibnamefont
  {Agapito}}, \bibinfo {author} {\bibfnamefont {N.}~\bibnamefont {Kioussis}},
  \bibinfo {author} {\bibfnamefont {W.~A.}\ \bibnamefont {Goddard}}, \ and\
  \bibinfo {author} {\bibfnamefont {N.~P.}\ \bibnamefont {Ong}},\ }\href
  {https://doi.org/10.1103/physrevlett.110.176401} {\bibfield  {journal}
  {\bibinfo  {journal} {Phys. Rev. Lett.}\ }\textbf {\bibinfo {volume} {110}}
  (\bibinfo {year} {2013})}\BibitemShut {NoStop}%
\bibitem [{\citenamefont {Winterfeld}\ \emph {et~al.}(2013)\citenamefont
  {Winterfeld}, \citenamefont {Agapito}, \citenamefont {Li}, \citenamefont
  {Kioussis}, \citenamefont {Blaha},\ and\ \citenamefont
  {Chen}}]{2013-PRB-Chen-strain}%
  \BibitemOpen
  \bibfield  {author} {\bibinfo {author} {\bibfnamefont {L.}~\bibnamefont
  {Winterfeld}}, \bibinfo {author} {\bibfnamefont {L.~A.}\ \bibnamefont
  {Agapito}}, \bibinfo {author} {\bibfnamefont {J.}~\bibnamefont {Li}},
  \bibinfo {author} {\bibfnamefont {N.}~\bibnamefont {Kioussis}}, \bibinfo
  {author} {\bibfnamefont {P.}~\bibnamefont {Blaha}}, \ and\ \bibinfo {author}
  {\bibfnamefont {Y.~P.}\ \bibnamefont {Chen}},\ }\href
  {https://doi.org/10.1103/physrevb.87.075143} {\bibfield  {journal} {\bibinfo
  {journal} {Phys. Rev. B}\ }\textbf {\bibinfo {volume} {87}} (\bibinfo {year}
  {2013})}\BibitemShut {NoStop}%
\bibitem [{\citenamefont {Barone}\ \emph {et~al.}(2013)\citenamefont {Barone},
  \citenamefont {Rauch}, \citenamefont {Sante}, \citenamefont {Henk},
  \citenamefont {Mertig},\ and\ \citenamefont
  {Picozzi}}]{2013-PRB-Barone-pressure}%
  \BibitemOpen
  \bibfield  {author} {\bibinfo {author} {\bibfnamefont {P.}~\bibnamefont
  {Barone}}, \bibinfo {author} {\bibfnamefont {T.}~\bibnamefont {Rauch}},
  \bibinfo {author} {\bibfnamefont {D.~D.}\ \bibnamefont {Sante}}, \bibinfo
  {author} {\bibfnamefont {J.}~\bibnamefont {Henk}}, \bibinfo {author}
  {\bibfnamefont {I.}~\bibnamefont {Mertig}}, \ and\ \bibinfo {author}
  {\bibfnamefont {S.}~\bibnamefont {Picozzi}},\ }\href
  {https://doi.org/10.1103/physrevb.88.045207} {\bibfield  {journal} {\bibinfo
  {journal} {Phys. Rev. B}\ }\textbf {\bibinfo {volume} {88}} (\bibinfo {year}
  {2013})}\BibitemShut {NoStop}%
\bibitem [{\citenamefont {Zhu}\ \emph {et~al.}(2012)\citenamefont {Zhu},
  \citenamefont {Cheng},\ and\ \citenamefont
  {Schwingenschl\"{o}gl}}]{2012-PRL-KAUST-pressure-layered}%
  \BibitemOpen
  \bibfield  {author} {\bibinfo {author} {\bibfnamefont {Z.}~\bibnamefont
  {Zhu}}, \bibinfo {author} {\bibfnamefont {Y.}~\bibnamefont {Cheng}}, \ and\
  \bibinfo {author} {\bibfnamefont {U.}~\bibnamefont {Schwingenschl\"{o}gl}},\
  }\href {https://doi.org/10.1103/physrevlett.108.266805} {\bibfield  {journal}
  {\bibinfo  {journal} {Phys. Rev. Lett.}\ }\textbf {\bibinfo {volume} {108}}
  (\bibinfo {year} {2012})}\BibitemShut {NoStop}%
\bibitem [{\citenamefont {Xi}\ \emph {et~al.}(2013)\citenamefont {Xi},
  \citenamefont {Ma}, \citenamefont {Liu}, \citenamefont {Chen}, \citenamefont
  {Ku}, \citenamefont {Berger}, \citenamefont {Martin}, \citenamefont
  {Tanner},\ and\ \citenamefont {Carr}}]{2013-PRL-Carr-TQPT}%
  \BibitemOpen
  \bibfield  {author} {\bibinfo {author} {\bibfnamefont {X.}~\bibnamefont
  {Xi}}, \bibinfo {author} {\bibfnamefont {C.}~\bibnamefont {Ma}}, \bibinfo
  {author} {\bibfnamefont {Z.}~\bibnamefont {Liu}}, \bibinfo {author}
  {\bibfnamefont {Z.}~\bibnamefont {Chen}}, \bibinfo {author} {\bibfnamefont
  {W.}~\bibnamefont {Ku}}, \bibinfo {author} {\bibfnamefont {H.}~\bibnamefont
  {Berger}}, \bibinfo {author} {\bibfnamefont {C.}~\bibnamefont {Martin}},
  \bibinfo {author} {\bibfnamefont {D.~B.}\ \bibnamefont {Tanner}}, \ and\
  \bibinfo {author} {\bibfnamefont {G.~L.}\ \bibnamefont {Carr}},\ }\href
  {http://dx.doi.org/10.1103/physrevlett.111.155701} {\bibfield  {journal}
  {\bibinfo  {journal} {Phys. Rev. Lett.}\ }\textbf {\bibinfo {volume} {111}}
  (\bibinfo {year} {2013})}\BibitemShut {NoStop}%
\bibitem [{\citenamefont {Xi}\ \emph {et~al.}(2014)\citenamefont {Xi},
  \citenamefont {He}, \citenamefont {Guan}, \citenamefont {Liu}, \citenamefont
  {Zhong}, \citenamefont {Schneeloch}, \citenamefont {Liu}, \citenamefont {Gu},
  \citenamefont {Du}, \citenamefont {Chen}, \citenamefont {Hong}, \citenamefont
  {Ku},\ and\ \citenamefont {Carr}}]{2014-PRL-Carr-bulk-signatures}%
  \BibitemOpen
  \bibfield  {author} {\bibinfo {author} {\bibfnamefont {X.}~\bibnamefont
  {Xi}}, \bibinfo {author} {\bibfnamefont {X.-G.}\ \bibnamefont {He}}, \bibinfo
  {author} {\bibfnamefont {F.}~\bibnamefont {Guan}}, \bibinfo {author}
  {\bibfnamefont {Z.}~\bibnamefont {Liu}}, \bibinfo {author} {\bibfnamefont
  {R.~D.}\ \bibnamefont {Zhong}}, \bibinfo {author} {\bibfnamefont {J.~A.}\
  \bibnamefont {Schneeloch}}, \bibinfo {author} {\bibfnamefont {T.~S.}\
  \bibnamefont {Liu}}, \bibinfo {author} {\bibfnamefont {G.~D.}\ \bibnamefont
  {Gu}}, \bibinfo {author} {\bibfnamefont {X.}~\bibnamefont {Du}}, \bibinfo
  {author} {\bibfnamefont {Z.}~\bibnamefont {Chen}}, \bibinfo {author}
  {\bibfnamefont {X.~G.}\ \bibnamefont {Hong}}, \bibinfo {author}
  {\bibfnamefont {W.}~\bibnamefont {Ku}}, \ and\ \bibinfo {author}
  {\bibfnamefont {G.~L.}\ \bibnamefont {Carr}},\ }\href
  {https://doi.org/10.1103/physrevlett.113.096401} {\bibfield  {journal}
  {\bibinfo  {journal} {Phys. Rev. Lett.}\ }\textbf {\bibinfo {volume} {113}}
  (\bibinfo {year} {2014})}\BibitemShut {NoStop}%
\bibitem [{\citenamefont {Sklyadneva}\ \emph {et~al.}(2016)\citenamefont
  {Sklyadneva}, \citenamefont {Rusinov}, \citenamefont {Heid}, \citenamefont
  {Bohnen}, \citenamefont {Echenique},\ and\ \citenamefont
  {Chulkov}}]{2016scireports-Sklydneva-KNa2Bi}%
  \BibitemOpen
  \bibfield  {author} {\bibinfo {author} {\bibfnamefont {I.~Y.}\ \bibnamefont
  {Sklyadneva}}, \bibinfo {author} {\bibfnamefont {I.~P.}\ \bibnamefont
  {Rusinov}}, \bibinfo {author} {\bibfnamefont {R.}~\bibnamefont {Heid}},
  \bibinfo {author} {\bibfnamefont {K.-P.}\ \bibnamefont {Bohnen}}, \bibinfo
  {author} {\bibfnamefont {P.~M.}\ \bibnamefont {Echenique}}, \ and\ \bibinfo
  {author} {\bibfnamefont {E.~V.}\ \bibnamefont {Chulkov}},\ }\href
  {http://dx.doi.org/10.1038/srep24137} {\bibfield  {journal} {\bibinfo
  {journal} {Sci. Rep.}\ }\textbf {\bibinfo {volume} {6}},\ \bibinfo {pages}
  {24137} (\bibinfo {year} {2016})}\BibitemShut {NoStop}%
\bibitem [{\citenamefont {Ohmura}\ \emph {et~al.}(2017)\citenamefont {Ohmura},
  \citenamefont {Higuchi}, \citenamefont {Ochiai}, \citenamefont {Kanou},
  \citenamefont {Ishikawa}, \citenamefont {Nakano}, \citenamefont {Nakayama},
  \citenamefont {Yamada},\ and\ \citenamefont
  {Sasagawa}}]{2017-PRB-japan-BiTeBr}%
  \BibitemOpen
  \bibfield  {author} {\bibinfo {author} {\bibfnamefont {A.}~\bibnamefont
  {Ohmura}}, \bibinfo {author} {\bibfnamefont {Y.}~\bibnamefont {Higuchi}},
  \bibinfo {author} {\bibfnamefont {T.}~\bibnamefont {Ochiai}}, \bibinfo
  {author} {\bibfnamefont {M.}~\bibnamefont {Kanou}}, \bibinfo {author}
  {\bibfnamefont {F.}~\bibnamefont {Ishikawa}}, \bibinfo {author}
  {\bibfnamefont {S.}~\bibnamefont {Nakano}}, \bibinfo {author} {\bibfnamefont
  {A.}~\bibnamefont {Nakayama}}, \bibinfo {author} {\bibfnamefont
  {Y.}~\bibnamefont {Yamada}}, \ and\ \bibinfo {author} {\bibfnamefont
  {T.}~\bibnamefont {Sasagawa}},\ }\href
  {https://doi.org/10.1103/physrevb.95.125203} {\bibfield  {journal} {\bibinfo
  {journal} {Phys. Rev. B}\ }\textbf {\bibinfo {volume} {95}} (\bibinfo {year}
  {2017})}\BibitemShut {NoStop}%
\bibitem [{\citenamefont {Park}\ \emph {et~al.}(2015)\citenamefont {Park},
  \citenamefont {Jin}, \citenamefont {Jo}, \citenamefont {Choi}, \citenamefont
  {Kang}, \citenamefont {Kampert}, \citenamefont {Rhyee}, \citenamefont {Jhi},\
  and\ \citenamefont {Kim}}]{2015Scireports-Park-BiTeI}%
  \BibitemOpen
  \bibfield  {author} {\bibinfo {author} {\bibfnamefont {J.}~\bibnamefont
  {Park}}, \bibinfo {author} {\bibfnamefont {K.-H.}\ \bibnamefont {Jin}},
  \bibinfo {author} {\bibfnamefont {Y.~J.}\ \bibnamefont {Jo}}, \bibinfo
  {author} {\bibfnamefont {E.~S.}\ \bibnamefont {Choi}}, \bibinfo {author}
  {\bibfnamefont {W.}~\bibnamefont {Kang}}, \bibinfo {author} {\bibfnamefont
  {E.}~\bibnamefont {Kampert}}, \bibinfo {author} {\bibfnamefont {J.-S.}\
  \bibnamefont {Rhyee}}, \bibinfo {author} {\bibfnamefont {S.-H.}\ \bibnamefont
  {Jhi}}, \ and\ \bibinfo {author} {\bibfnamefont {J.~S.}\ \bibnamefont
  {Kim}},\ }\href {http://dx.doi.org/10.1038/srep15973} {\bibfield  {journal}
  {\bibinfo  {journal} {Sci. Rep.}\ }\textbf {\bibinfo {volume} {5}},\ \bibinfo
  {pages} {15973} (\bibinfo {year} {2015})}\BibitemShut {NoStop}%
\bibitem [{\citenamefont {Sun}\ \emph {et~al.}(2016)\citenamefont {Sun},
  \citenamefont {Wang}, \citenamefont {Wu}, \citenamefont {Felser},
  \citenamefont {Liu},\ and\ \citenamefont {Yan}}]{2016-PRB-Sun-pressure}%
  \BibitemOpen
  \bibfield  {author} {\bibinfo {author} {\bibfnamefont {Y.}~\bibnamefont
  {Sun}}, \bibinfo {author} {\bibfnamefont {Q.-Z.}\ \bibnamefont {Wang}},
  \bibinfo {author} {\bibfnamefont {S.-C.}\ \bibnamefont {Wu}}, \bibinfo
  {author} {\bibfnamefont {C.}~\bibnamefont {Felser}}, \bibinfo {author}
  {\bibfnamefont {C.-X.}\ \bibnamefont {Liu}}, \ and\ \bibinfo {author}
  {\bibfnamefont {B.}~\bibnamefont {Yan}},\ }\href
  {https://doi.org/10.1103/physrevb.93.205303} {\bibfield  {journal} {\bibinfo
  {journal} {Phys. Rev. B}\ }\textbf {\bibinfo {volume} {93}} (\bibinfo {year}
  {2016})}\BibitemShut {NoStop}%
\bibitem [{\citenamefont {Zhu}\ \emph {et~al.}(2016)\citenamefont {Zhu},
  \citenamefont {Li},\ and\ \citenamefont {Li}}]{2016-PRB-Zintl}%
  \BibitemOpen
  \bibfield  {author} {\bibinfo {author} {\bibfnamefont {Z.}~\bibnamefont
  {Zhu}}, \bibinfo {author} {\bibfnamefont {M.}~\bibnamefont {Li}}, \ and\
  \bibinfo {author} {\bibfnamefont {J.}~\bibnamefont {Li}},\ }\href
  {https://doi.org/10.1103/physrevb.94.155121} {\bibfield  {journal} {\bibinfo
  {journal} {Phys. Rev. B}\ }\textbf {\bibinfo {volume} {94}} (\bibinfo {year}
  {2016})}\BibitemShut {NoStop}%
\bibitem [{\citenamefont {Zhao}\ \emph {et~al.}(2015)\citenamefont {Zhao},
  \citenamefont {Wang}, \citenamefont {Gu},\ and\ \citenamefont
  {Duan}}]{2015-PRB-Zhao-TCI}%
  \BibitemOpen
  \bibfield  {author} {\bibinfo {author} {\bibfnamefont {L.}~\bibnamefont
  {Zhao}}, \bibinfo {author} {\bibfnamefont {J.}~\bibnamefont {Wang}}, \bibinfo
  {author} {\bibfnamefont {B.-L.}\ \bibnamefont {Gu}}, \ and\ \bibinfo {author}
  {\bibfnamefont {W.}~\bibnamefont {Duan}},\ }\href
  {https://doi.org/10.1103/physrevb.91.195320} {\bibfield  {journal} {\bibinfo
  {journal} {Phys. Rev. B}\ }\textbf {\bibinfo {volume} {91}} (\bibinfo {year}
  {2015})}\BibitemShut {NoStop}%
\bibitem [{\citenamefont {Kohn}\ and\ \citenamefont
  {Sham}(1965)}]{1965-PhysRev-Kohn-Vxc}%
  \BibitemOpen
  \bibfield  {author} {\bibinfo {author} {\bibfnamefont {W.}~\bibnamefont
  {Kohn}}\ and\ \bibinfo {author} {\bibfnamefont {L.~J.}\ \bibnamefont
  {Sham}},\ }\href {\doibase 10.1103/PhysRev.140.A1133} {\bibfield  {journal}
  {\bibinfo  {journal} {Phys. Rev.}\ }\textbf {\bibinfo {volume} {140}},\
  \bibinfo {pages} {A1133} (\bibinfo {year} {1965})}\BibitemShut {NoStop}%
\bibitem [{\citenamefont {Kresse}\ and\ \citenamefont
  {Furthm{\"u}ller}(1996)}]{1966CMS-Kresse-Ecalcs}%
  \BibitemOpen
  \bibfield  {author} {\bibinfo {author} {\bibfnamefont {G.}~\bibnamefont
  {Kresse}}\ and\ \bibinfo {author} {\bibfnamefont {J.}~\bibnamefont
  {Furthm{\"u}ller}},\ }\href {\doibase 10.1016/0927-0256(96)00008-0}
  {\bibfield  {journal} {\bibinfo  {journal} {Comput. Mater. Sci.}\ }\textbf
  {\bibinfo {volume} {6}},\ \bibinfo {pages} {15} (\bibinfo {year}
  {1996})}\BibitemShut {NoStop}%
\bibitem [{\citenamefont {Kresse}\ and\ \citenamefont
  {Furthm\"uller}(1996)}]{1996PRB-Kresse-iterativeEcalcs}%
  \BibitemOpen
  \bibfield  {author} {\bibinfo {author} {\bibfnamefont {G.}~\bibnamefont
  {Kresse}}\ and\ \bibinfo {author} {\bibfnamefont {J.}~\bibnamefont
  {Furthm\"uller}},\ }\href {\doibase 10.1103/PhysRevB.54.11169} {\bibfield
  {journal} {\bibinfo  {journal} {Phys. Rev. B}\ }\textbf {\bibinfo {volume}
  {54}},\ \bibinfo {pages} {11169} (\bibinfo {year} {1996})}\BibitemShut
  {NoStop}%
\bibitem [{\citenamefont {Bl{\"o}chl}(1994)}]{1994PRB-Blochl-PAW}%
  \BibitemOpen
  \bibfield  {author} {\bibinfo {author} {\bibfnamefont {P.~E.}\ \bibnamefont
  {Bl{\"o}chl}},\ }\href {\doibase 10.1103/PhysRevB.50.17953} {\bibfield
  {journal} {\bibinfo  {journal} {Phys. Rev. B}\ }\textbf {\bibinfo {volume}
  {50}},\ \bibinfo {pages} {17953} (\bibinfo {year} {1994})}\BibitemShut
  {NoStop}%
\bibitem [{\citenamefont {Kresse}\ and\ \citenamefont
  {Joubert}(1999)}]{1999PRB-Kresse-PAW}%
  \BibitemOpen
  \bibfield  {author} {\bibinfo {author} {\bibfnamefont {G.}~\bibnamefont
  {Kresse}}\ and\ \bibinfo {author} {\bibfnamefont {D.}~\bibnamefont
  {Joubert}},\ }\href {\doibase 10.1103/PhysRevB.59.1758} {\bibfield  {journal}
  {\bibinfo  {journal} {Phys. Rev. B}\ }\textbf {\bibinfo {volume} {59}},\
  \bibinfo {pages} {1758} (\bibinfo {year} {1999})}\BibitemShut {NoStop}%
\bibitem [{\citenamefont {Heyd}\ \emph {et~al.}(2003)\citenamefont {Heyd},
  \citenamefont {Scuseria},\ and\ \citenamefont {Ernzerhof}}]{HSE-2003}%
  \BibitemOpen
  \bibfield  {author} {\bibinfo {author} {\bibfnamefont {J.}~\bibnamefont
  {Heyd}}, \bibinfo {author} {\bibfnamefont {G.~E.}\ \bibnamefont {Scuseria}},
  \ and\ \bibinfo {author} {\bibfnamefont {M.}~\bibnamefont {Ernzerhof}},\
  }\href {\doibase 10.1063/1.1564060} {\bibfield  {journal} {\bibinfo
  {journal} {J Chem. Phys.}\ }\textbf {\bibinfo {volume} {118}},\ \bibinfo
  {pages} {8207} (\bibinfo {year} {2003})}\BibitemShut {NoStop}%
\bibitem [{\citenamefont {Heyd}\ \emph {et~al.}(2006)\citenamefont {Heyd},
  \citenamefont {Scuseria},\ and\ \citenamefont {Ernzerhof}}]{HSE-2006}%
  \BibitemOpen
  \bibfield  {author} {\bibinfo {author} {\bibfnamefont {J.}~\bibnamefont
  {Heyd}}, \bibinfo {author} {\bibfnamefont {G.~E.}\ \bibnamefont {Scuseria}},
  \ and\ \bibinfo {author} {\bibfnamefont {M.}~\bibnamefont {Ernzerhof}},\
  }\href {\doibase 10.1063/1.2204597} {\bibfield  {journal} {\bibinfo
  {journal} {J Chem. Phys.}\ }\textbf {\bibinfo {volume} {124}},\ \bibinfo
  {pages} {219906} (\bibinfo {year} {2006})}\BibitemShut {NoStop}%
\bibitem [{\citenamefont {Monkhorst}\ and\ \citenamefont
  {Pack}(1976)}]{1976PRB-Monkhorst-BZintegration}%
  \BibitemOpen
  \bibfield  {author} {\bibinfo {author} {\bibfnamefont {H.~J.}\ \bibnamefont
  {Monkhorst}}\ and\ \bibinfo {author} {\bibfnamefont {J.~D.}\ \bibnamefont
  {Pack}},\ }\href {\doibase 10.1103/PhysRevB.13.5188} {\bibfield  {journal}
  {\bibinfo  {journal} {Phys. Rev. B}\ }\textbf {\bibinfo {volume} {13}},\
  \bibinfo {pages} {5188} (\bibinfo {year} {1976})}\BibitemShut {NoStop}%
\bibitem [{\citenamefont {Soluyanov}\ and\ \citenamefont
  {Vanderbilt}(2011)}]{2011-PRB-Soluyanov-Z2}%
  \BibitemOpen
  \bibfield  {author} {\bibinfo {author} {\bibfnamefont {A.~A.}\ \bibnamefont
  {Soluyanov}}\ and\ \bibinfo {author} {\bibfnamefont {D.}~\bibnamefont
  {Vanderbilt}},\ }\href {http://dx.doi.org/10.1103/physrevb.83.235401}
  {\bibfield  {journal} {\bibinfo  {journal} {Phys. Rev. B}\ }\textbf {\bibinfo
  {volume} {83}} (\bibinfo {year} {2011})}\BibitemShut {NoStop}%
\bibitem [{\citenamefont {Fu}\ and\ \citenamefont
  {Kane}(2006)}]{2006-PRB-Kane-Z2}%
  \BibitemOpen
  \bibfield  {author} {\bibinfo {author} {\bibfnamefont {L.}~\bibnamefont
  {Fu}}\ and\ \bibinfo {author} {\bibfnamefont {C.~L.}\ \bibnamefont {Kane}},\
  }\href {http://dx.doi.org/10.1103/physrevb.74.195312} {\bibfield  {journal}
  {\bibinfo  {journal} {Phys. Rev. B}\ }\textbf {\bibinfo {volume} {74}}
  (\bibinfo {year} {2006})}\BibitemShut {NoStop}%
\bibitem [{\citenamefont {Yu}\ \emph {et~al.}(2011)\citenamefont {Yu},
  \citenamefont {Qi}, \citenamefont {Bernevig}, \citenamefont {Fang},\ and\
  \citenamefont {Dai}}]{2011-PRB-Bernevig-Z2}%
  \BibitemOpen
  \bibfield  {author} {\bibinfo {author} {\bibfnamefont {R.}~\bibnamefont
  {Yu}}, \bibinfo {author} {\bibfnamefont {X.~L.}\ \bibnamefont {Qi}}, \bibinfo
  {author} {\bibfnamefont {A.}~\bibnamefont {Bernevig}}, \bibinfo {author}
  {\bibfnamefont {Z.}~\bibnamefont {Fang}}, \ and\ \bibinfo {author}
  {\bibfnamefont {X.}~\bibnamefont {Dai}},\ }\href
  {http://dx.doi.org/10.1103/physrevb.84.075119} {\bibfield  {journal}
  {\bibinfo  {journal} {Phys. Rev. B}\ }\textbf {\bibinfo {volume} {84}}
  (\bibinfo {year} {2011})}\BibitemShut {NoStop}%
\bibitem [{\citenamefont {Mostofi}\ \emph {et~al.}(2014)\citenamefont
  {Mostofi}, \citenamefont {Yates}, \citenamefont {Pizzi}, \citenamefont {Lee},
  \citenamefont {Souza}, \citenamefont {Vanderbilt},\ and\ \citenamefont
  {Marzari}}]{2014ComputPhysCommun-Mostofi-Wannier90}%
  \BibitemOpen
  \bibfield  {author} {\bibinfo {author} {\bibfnamefont {A.~A.}\ \bibnamefont
  {Mostofi}}, \bibinfo {author} {\bibfnamefont {J.~R.}\ \bibnamefont {Yates}},
  \bibinfo {author} {\bibfnamefont {G.}~\bibnamefont {Pizzi}}, \bibinfo
  {author} {\bibfnamefont {Y.-S.}\ \bibnamefont {Lee}}, \bibinfo {author}
  {\bibfnamefont {I.}~\bibnamefont {Souza}}, \bibinfo {author} {\bibfnamefont
  {D.}~\bibnamefont {Vanderbilt}}, \ and\ \bibinfo {author} {\bibfnamefont
  {N.}~\bibnamefont {Marzari}},\ }\href {\doibase 10.1016/j.cpc.2014.05.003}
  {\bibfield  {journal} {\bibinfo  {journal} {Comput. Phys. Commun.}\ }\textbf
  {\bibinfo {volume} {185}},\ \bibinfo {pages} {2309} (\bibinfo {year}
  {2014})}\BibitemShut {NoStop}%
\bibitem [{\citenamefont {Wu}\ \emph {et~al.}(2017)\citenamefont {Wu},
  \citenamefont {Zhang}, \citenamefont {Song}, \citenamefont {Troyer},\ and\
  \citenamefont {Soluyanov}}]{opensource-QuanSheng-Wannier-tools}%
  \BibitemOpen
  \bibfield  {author} {\bibinfo {author} {\bibfnamefont {Q.}~\bibnamefont
  {Wu}}, \bibinfo {author} {\bibfnamefont {S.}~\bibnamefont {Zhang}}, \bibinfo
  {author} {\bibfnamefont {H.-F.}\ \bibnamefont {Song}}, \bibinfo {author}
  {\bibfnamefont {M.}~\bibnamefont {Troyer}}, \ and\ \bibinfo {author}
  {\bibfnamefont {A.~A.}\ \bibnamefont {Soluyanov}},\ }\href
  {http://arxiv.org/abs/1703.07789} {\  (\bibinfo {year} {2017})},\ \Eprint
  {http://arxiv.org/abs/1703.07789} {arXiv:1703.07789} \BibitemShut {NoStop}%
\bibitem [{\citenamefont {Teng{\aa}}\ \emph {et~al.}(2005)\citenamefont
  {Teng{\aa}}, \citenamefont {Garc{\'\i}a-Garc{\'\i}a}, \citenamefont
  {Mikhaylushkin}, \citenamefont {Espinosa-Arronte}, \citenamefont
  {Andersson},\ and\ \citenamefont {H{\"a}ussermann}}]{tengaa2005}%
  \BibitemOpen
  \bibfield  {author} {\bibinfo {author} {\bibfnamefont {A.}~\bibnamefont
  {Teng{\aa}}}, \bibinfo {author} {\bibfnamefont {F.~J.}\ \bibnamefont
  {Garc{\'\i}a-Garc{\'\i}a}}, \bibinfo {author} {\bibfnamefont {A.~S.}\
  \bibnamefont {Mikhaylushkin}}, \bibinfo {author} {\bibfnamefont
  {B.}~\bibnamefont {Espinosa-Arronte}}, \bibinfo {author} {\bibfnamefont
  {M.}~\bibnamefont {Andersson}}, \ and\ \bibinfo {author} {\bibfnamefont
  {U.}~\bibnamefont {H{\"a}ussermann}},\ }\href {\doibase
  10.1002/chin.200608009} {\bibfield  {journal} {\bibinfo  {journal} {Chem.
  Mater.}\ }\textbf {\bibinfo {volume} {17}},\ \bibinfo {pages} {6080}
  (\bibinfo {year} {2005})}\BibitemShut {NoStop}%
\bibitem [{\citenamefont {Shay}\ and\ \citenamefont
  {Wernick}(2013)}]{shay2013}%
  \BibitemOpen
  \bibfield  {author} {\bibinfo {author} {\bibfnamefont {J.~L.}\ \bibnamefont
  {Shay}}\ and\ \bibinfo {author} {\bibfnamefont {J.~H.}\ \bibnamefont
  {Wernick}},\ }\href@noop {} {\emph {\bibinfo {title} {Ternary Chalcopyrite
  Semiconductors: Growth, Electronic Properties, and Applications:
  International Series of Monographs in The Science of The Solid State}}},\
  Vol.~\bibinfo {volume} {7}\ (\bibinfo  {publisher} {Elsevier},\ \bibinfo
  {year} {2013})\BibitemShut {NoStop}%
\bibitem [{\citenamefont {Feng}\ \emph {et~al.}(2011)\citenamefont {Feng},
  \citenamefont {Xiao}, \citenamefont {Ding},\ and\ \citenamefont
  {Yao}}]{2011-feng-prl-controversial}%
  \BibitemOpen
  \bibfield  {author} {\bibinfo {author} {\bibfnamefont {W.}~\bibnamefont
  {Feng}}, \bibinfo {author} {\bibfnamefont {D.}~\bibnamefont {Xiao}}, \bibinfo
  {author} {\bibfnamefont {J.}~\bibnamefont {Ding}}, \ and\ \bibinfo {author}
  {\bibfnamefont {Y.}~\bibnamefont {Yao}},\ }\href
  {https://doi.org/10.1103/physrevlett.106.016402} {\bibfield  {journal}
  {\bibinfo  {journal} {Phys. Rev. Lett.}\ }\textbf {\bibinfo {volume} {106}}
  (\bibinfo {year} {2011})}\BibitemShut {NoStop}%
\bibitem [{\citenamefont {Feng}\ and\ \citenamefont
  {Yao}(2012)}]{2012-Feng-Chinese-Review}%
  \BibitemOpen
  \bibfield  {author} {\bibinfo {author} {\bibfnamefont {W.}~\bibnamefont
  {Feng}}\ and\ \bibinfo {author} {\bibfnamefont {Y.}~\bibnamefont {Yao}},\
  }\href {\doibase 10.1007/s11433-012-4929-9} {\bibfield  {journal} {\bibinfo
  {journal} {Sci. China Phys., Mech. and Astron.}\ }\textbf {\bibinfo {volume}
  {55}},\ \bibinfo {pages} {2199} (\bibinfo {year} {2012})}\BibitemShut
  {NoStop}%
\bibitem [{\citenamefont {Luttinger}(1956)}]{luttinger-original}%
  \BibitemOpen
  \bibfield  {author} {\bibinfo {author} {\bibfnamefont {J.~M.}\ \bibnamefont
  {Luttinger}},\ }\href {\doibase 10.1103/PhysRev.102.1030} {\bibfield
  {journal} {\bibinfo  {journal} {Phys. Rev.}\ }\textbf {\bibinfo {volume}
  {102}},\ \bibinfo {pages} {1030} (\bibinfo {year} {1956})}\BibitemShut
  {NoStop}%
\end{thebibliography}
%

\end{document}